\documentclass[10pt]{article} 
\usepackage[preprint]{tmlr}

\usepackage{amsmath,amsfonts,bm}

\def\eqref#1{equation~\ref{#1}}

\def\1{\bm{1}}

\DeclareMathAlphabet{\mathsfit}{\encodingdefault}{\sfdefault}{m}{sl}
\SetMathAlphabet{\mathsfit}{bold}{\encodingdefault}{\sfdefault}{bx}{n}

\usepackage{hyperref}
\usepackage{url}
\usepackage{enumitem}
\usepackage{cleveref}
\usepackage{float}
\usepackage{array}
\usepackage{makecell}  
\usepackage{booktabs}  
\newcolumntype{P}[1]{>{\raggedright\arraybackslash}p{#1}}   
\usepackage[T1]{fontenc} 
\usepackage{hyperref}
\usepackage{tikz}
\usepackage{xcolor}
\usepackage{arydshln}
\usepackage{pifont}
\usepackage{multirow}
\usepackage{supertabular}
\usepackage[most]{tcolorbox}

\newcommand{\cmark}{\ding{51}} 
\newcommand{\xmark}{\ding{55}} 

\title{Legal Retrieval for Public Defenders}

\author{\name Dominik Stammbach \email dominsta@princeton.edu \\
      \addr Princeton University
      \AND
      \name Kylie Zhang \email kylie.zhang@princeton.edu \\
      \addr Princeton University
      \AND
      \name Patty Liu \email patty.liu@princeton.edu \\
      \addr Princeton University
      \AND
      \name Nimra Nadeem \email nnadeem@princeton.edu \\
      \addr Princeton University
      \AND
      \name Inyoung Cheong \email iycheong@princeton.edu \\
      \addr Princeton University
      \AND
      \name Lucia Zheng \email zlucia@stanford.edu \\
      \addr Stanford University
      \AND
      \name Peter Henderson \email peter.henderson@princeton.edu \\
      \addr Princeton University
}

\def\month{08}  
\def\year{2026} 
\def\openreview{\url{https://openreview.net/forum?id=HnbKQGRnDt}} 

\begin{document}

\maketitle

\noindent\href{https://huggingface.co/datasets/ai-law-society-lab/PublicDefenseDataset}{Dataset} \\
\href{https://github.com/princeton-polaris-lab/PublicDefenderRetrieval}{Replication Package}

\begin{abstract}
AI tools are suggested as solutions to assist public agencies with heavy workloads. In public defense---where a constitutional right to counsel meets the complexities of law, overwhelming caseloads, and constrained resources---practitioners face especially taxing conditions. Yet, there is little evidence of how AI could meaningfully support defenders' day-to-day work. In partnership with the New Jersey Office of the Public Defender, we develop the OPD Resource Library, a retrieval tool which surfaces relevant appellate briefs to streamline legal research and writing.
We show that existing retrieval benchmarks fail to transfer to real public defense research, however adding domain knowledge improves retrieval quality. This includes query expansion with legal reasoning, domain-specific data and curated synthetic examples. To facilitate further research, we release a taxonomy of realistic defender search queries and a manually annotated evaluation dataset for public defense retrieval. This benchmark is highly correlated with a proprietary retrieval dataset annotated by experienced public defenders. Our work improves on the status quo of realistic legal retrieval benchmarking and illustrates one approach to applying AI in a real-world public interest setting.
\end{abstract}

\section{Introduction}

In the United States, individuals facing criminal charges have a right to counsel. This right is guaranteed by the Sixth Amendment and reaffirmed in \citet{justia_gideon_v._wainwright}. For those unable to afford private counsel, representation is provided by public defenders. In practice however, defenders often face severe resource constraints and overwhelming caseloads \citep{Pace2023}, while having to navigate the complexities of today's legal system. Combined, these can undermine promises of fair and equal legal representation for clients relying on public defense.

Advances in natural language processing, particularly in foundation models, have raised hopes that AI tools could assist public defenders by streamlining time-consuming tasks, including for example legal research or drafting of briefs \citep{bommasani2022opportunitiesrisksfoundationmodels,mahari-etal-2023-law, cheong2025aiaugmentaccessjustice}. However, despite rapid advances in model capabilities and legal benchmark performance \citep{10.5555/3666122.3668037, dominguezolmedo2025lawmapowerspecializationlegal}, there remain very few examples of concrete real-world implementation and evaluation of AI within public defender offices. 

This gap limits our understanding of what types of AI use cases are feasible, safe, and can genuinely empower defenders in day-to-day legal practice. Public defense is a high-stakes setting where errors can directly affect the outcome of cases and clients.
Hence, carefully balancing trade-offs between accuracy, reliability and risks of failure become essential design constraints in such applications. Consequently, any AI assistance for public defenders must prioritize verifiable and trustworthy outputs.

In this work, we partner with the New Jersey Office of the Public Defender (NJOPD) to identify, develop, and evaluate such an AI use case. Defenders often specialize in specific legal areas, like felonies or misdemeanors. If handed a case outside their area of specialization, they often consult colleagues to obtain past briefs within the office handling similar cases. Such past briefs, especially those from the appellate section written by experienced defenders, offer overviews, reusable legal arguments, applicable precedent and overall guidance for how to navigate similar legal circumstances.

Inspired by such current office practices, we developed the OPD Resource Library. The Library leverages foundation model-based embeddings for retrieval, along other components, to search over all appellate briefs within the office and surfaces relevant ones. By providing access to relevant briefs, defenders can reuse legal arguments, precedent and other applicable information. Thus, the Library is designed to streamline brief drafting, but also maintains more consistency within the office: defenders may be more likely to leverage already-identified winning strategies and follow best practices.

During multiple evaluation rounds, experienced defenders submitted realistic search queries and assessed the quality of the tool. The submitted queries are highly diverse: They include broad topical searches, specific legal arguments, pinpoint citations, statutory definitions and doctrinal status checks. These queries reflect the heterogeneous and practice-driven information needs of public defenders. 

\begin{figure}[t!]
    \centering
    \begin{minipage}[t]{0.48\linewidth}
        \centering
        \includegraphics[width=\linewidth]{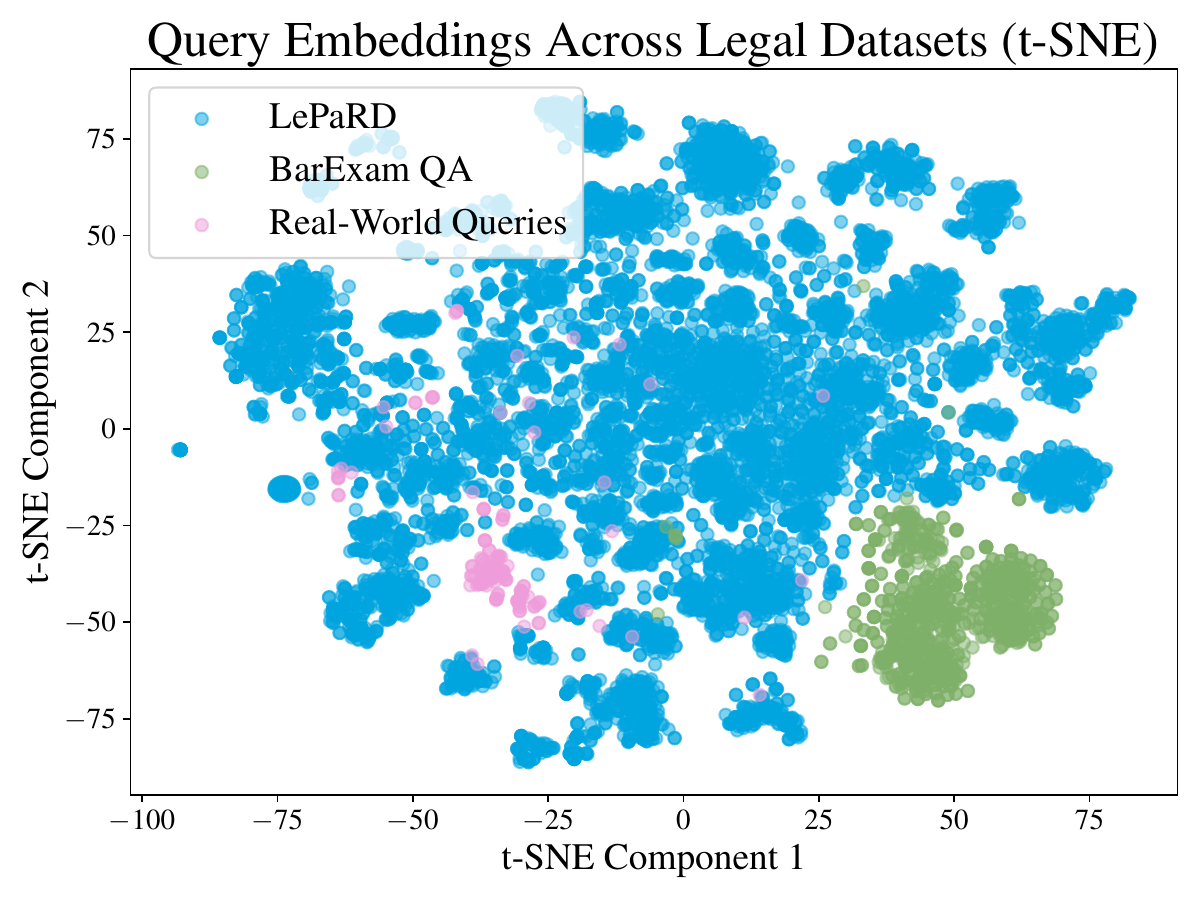}
    \end{minipage}
    \hfill
    \begin{minipage}[t]{0.48\linewidth}
        \centering
        \includegraphics[width=\linewidth]{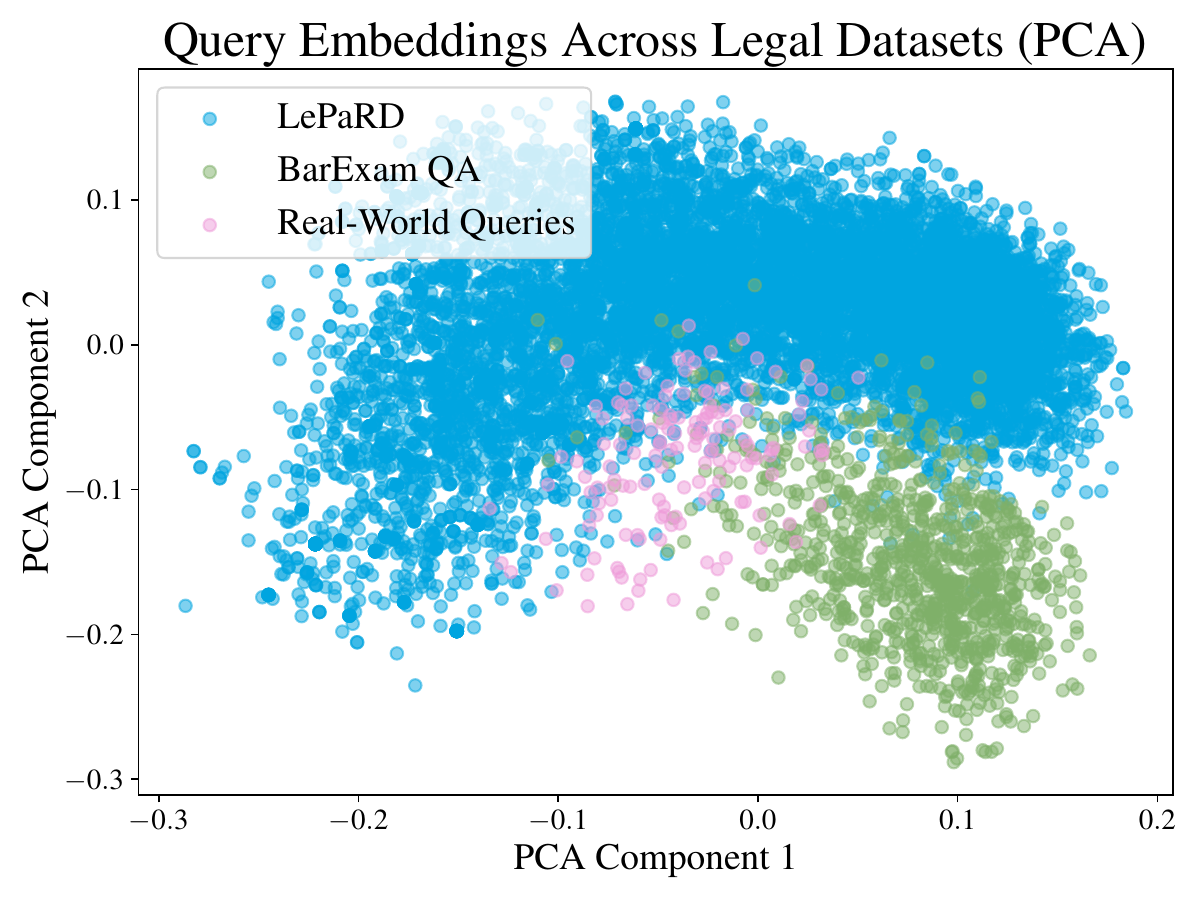}
    \end{minipage}
    \caption{Visualizations of defender queries (pink dots) and other queries in two legal search datasets: LePaRD \citep{mahari-etal-2024-lepard} and BarExam-QA \citep{barexam_qa}. We compute low-dimensional projections with t-SNE \citep{vandermaaten08a} and PCA \citep{sklearn}. We observe that queries from different datasets are separable in the embedding spaces.}    
    \label{fig:visualization_embeddings}
    
\end{figure}

In contrast, existing legal retrieval datasets are often constructed somewhat artificially, e.g., by reusing bar-exam questions as queries \citep{barexam_qa}, finding similar cases given a query case \citep{coliee_2022_summary},
or by heuristically linking preceding context to quoted legal passages \citep{hou-etal-2025-clerc, mahari-etal-2024-lepard}. In Appendix Table \ref{tab:legal_retrieval_datasets}, we show that no existing U.S. legal retrieval datasets contain both realistic queries and manually verified target paragraphs. While existing datasets are valuable for controlled benchmarking and measuring progress on legal retrieval, they poorly approximate more realistic contexts. Figure \ref{fig:visualization_embeddings} visualizes query embeddings from BarExam-QA \citep{barexam_qa}, LePaRD \citep{mahari-etal-2024-lepard}, and public defender queries, which are separable in embedding space. In Section \ref{sec:experiments}, we will show that training on such datasets decreases performance on public defense retrieval.

On the other hand, we improve recall by (1) using more recent and larger embedding models \citep{lee2024nv, zhang2025qwen3embeddingadvancingtext} and (2) leveraging domain knowledge. To introduce domain knowledge, we generate curated synthetic training data using a fine-tuned defender query generation model, and filter that data with a fine-tuned legal reranker \citep[similar to][]{promptagator}.
We further expand queries in this dataset using the IRAC framework\footnote{Issue, Rule, Application, and Conclusion.} developed for legal analysis. See Section \ref{sec:experiments} for full results.

Our work is a first step toward practical AI for public defense, inspired by current office practices. \citet{ott2022mapping} argue that future benchmarks should emphasize real-world utility. In our work, we offer a step towards more realistic legal NLP benchmarking, enabled by collaboratively developing the OPD Resource Library in partnership with the NJOPD. To stimulate further research on legal retrieval for public defenders, we release a manually annotated dataset, fine-tuned models and replication code.\footnote{replication package: \href{https://github.com/princeton-polaris-lab/PublicDefenderRetrieval}{https://github.com/princeton-polaris-lab/PublicDefenderRetrieval}}

Combined, these artifacts could help develop similar libraries for other defender offices or pro-bono clinics, but also measure progress in more realistic legal retrieval settings. More broadly, this collaboration illustrates how partnerships with public institutions allow situating NLP research in real-world applications. To summarize, we make the following contributions:

\begin{itemize}[nosep]
    \item We introduce the public defense retrieval task: retrieving relevant passages from existing appellate defense briefs. We construct an accompanying evaluation dataset comprised of 170 queries and 543 human-annotated relevant paragraphs from publicly available documents (Section \ref{sec:briefbank}).
    \item Evaluation of eight pre-trained retrieval models on public defense search. While larger models perform better, the best model only achieves 37.08\% Recall@5, demonstrating substantial room for improvement on this task (Table \ref{tab:zeroshot-retrieval}).
    \item A taxonomy of defender search queries, discussing shortcomings of current models and informing future work on public defense search (Section \ref{sec:taxonomy}).
    \item Evidence of a mismatch between existing legal retrieval benchmarks and real-world use cases. We find that fine-tuning retrieval models on existing legal retrieval benchmarks degrades performance, while legal domain adaptation, fine-tuning on carefully curated synthetic data and query expansion strategies improve performance (Section \ref{sec:experiments}).

\end{itemize}

\section{Public Defense Retrieval}\label{sec:AI_for_public_defenders}

\citet{cheong2025aiaugmentaccessjustice} group public defense work into five pillars, of which two seem especially suitable to AI assistance: evidence investigation and legal research and writing. For legal research and writing, they report that AI would be most useful to generate surveys of information, provide starting points, draft documents and narrow down case searches. 

This perspective is reinforced in our collaboration with the NJOPD. A well-written brief functions as a structured survey of a legal issue. The office encourages defenders to use such briefs as starting points if they have to handle a case outside their specialization. Thus, access to such briefs streamlines legal drafting through reuse of existing materials. Lastly, these briefs contain relevant legal precedent, and thus indirectly narrow down case search.

Apart from searching for briefs, we also explored other AI use cases to empower defenders. These include (1) directly answering legal queries using generative AI with a closed-source RAG tool (NotebookLM), and (2) searching through federal and state case law. However, both failed to deliver sufficient levels of accuracy, verifiability, and transparency, hence we define public defense retrieval as searching over past briefs, materials defenders already use and trust. Building accurate search over briefs also serves as a lens into the collective knowledge acquired within an office over time, and makes that knowledge accessible.

\subsection{Retrieval vs. Generation}

Past work pointed out potential usefulness of generative AI for legal work \citep{bommasani2022opportunitiesrisksfoundationmodels,mahari-etal-2023-law}, which has been supported by promising legal evaluation of LLMs, such as GPT-4 passing the bar exam \citep{10.1098/rsta.2023.0254}. \citet{SchwarczChoi2023AITools} find that LLMs help law students draft legal documents faster. Given such evidence, the office explored the potential of LLMs to assist public defenders in their day-to-day work. Experienced defenders submitted 100 queries to NotebookLM, a Retrieval-Augmented Generation (RAG) application. Although NotebookLM had access to a small set of relevant briefs, 66\% of the NotebookLM generations contained issues. The three main failure modes consisted of: 

\begin{enumerate}[nosep]
    \item Hallucinations (of e.g., citations) or incorrect references to source materials.
    \item Failure of the model to address nuanced legal contexts.
    \item Incomplete or verbose outputs, or generating unrelated information altogether.
\end{enumerate}

Strikingly, \citet{cheong2025aiaugmentaccessjustice} report that 85\% of public defenders they interviewed currently doubt AI can reliably verify research output, describing the same reasons (hallucinations, failure to handle nuanced legal context and incomplete output) NJOPD identified in their explorative evaluation. Given the high stakes of public defense, where hallucinated citations and misstated precedents can affect the outcome of cases, such failure rates are unacceptable in practice. Beyond accuracy, using commercial generative models raises confidentiality risks: case details submitted to proprietary APIs may fall outside attorney-client privilege and be subject to mandatory disclosure \citep{cheong2024not, privilege}. Generative models also provide limited transparency about sources, making it difficult for attorneys to verify output accuracy.

\subsection{Brief vs. Case Law Retrieval}
We also considered retrieval over all state and federal case law, which would directly return relevant precedent. However, this risks surfacing opinions that are no longer good law. As prior work shows, approximately 7.8\% of lower-court decisions are later reversed on appeal \citep{edwards2019affirmation}. Because defenders cannot rely on non-binding or outdated precedent, case law search alone without information about whether a case is still good law seems too unreliable for practical use cases. While commercial systems include such information, automatically detecting overturned case law remains an active and unresolved research problem \citep{zhang2025llmstrulyunderstandprecedent}. In the Resource Library, we address this by including internal documents and public directives, both containing, among others, overturned case alerts and up-to-date best practices for frequently occurring legal issues.

\subsection{Task Definition}

Formally, we define the public defense retrieval task as follows: Given a user query $q$, which may be related to a citation, rule reference, doctrinal question, or natural-language description of a legal concept, the goal is to retrieve the most relevant paragraphs $p_i$ from a corpus of prior briefs, other internal documents and public directives. Each query can have multiple relevant paragraphs.

\section{The OPD Resource Library and the PD Retrieval Dataset} \label{sec:briefbank}

In this section, we describe the Resource Library in more detail, and the associated PD dataset construction process.

\begin{table}[t!]
    \centering
    \caption{Legal Search Queries and Associated Paragraphs}
    \scriptsize
    \begin{tabular}{p{4cm} p{11.6cm}} \\
    \textbf{Search Query} & \textbf{Relevant Paragraph} \\ \toprule
      Difference between reasonable suspicion and probable cause & State v. Rodriguez, 172 N.J. 117, 126 (2002) (quoting Terry v. Ohio, 392 U.S. 1, 21 (1968)).  Although reasonable suspicion is a less demanding standard than probable cause, “[n]either ‘inarticulate hunches’ nor an arresting officer’s subjective good faith can justify infringement of a citizen’s constitutionally guaranteed rights. \\ 
    does the fruit of the poisonous tree doctrine apply in the fifth amendment context, too? & See also State v. O’Neill, 104 N.J. 148, 171, n.13 (2007) (“The fruit-of-the-poisonous-tree doctrine denies the prosecution the use of derivative evidence obtained as a result of a Fourth or Fifth Amendment violation. ”)  Moreover, New Jersey’s privilege against self-incrimination is so ingrained and deeply rooted in the State’s common law that it has always been considered as of “constitutional magnitude” offering “broader protection than its Fifth Amendment federal counterpart.  \\
    Is consent to search valid if the motor vehicle stop was illegal? & Because the police unlawfully prolonged the detention and sought consent to search without reasonable suspicion of criminal activity, in violation of the Fourth Amendment and Article I, paragraph 7 of the New Jersey Constitution, all of the evidence must be suppressed.  U.S. Const. amends.  IV, XIV; N.J. Const. art.  1, para. 7; Wong Sun v. United States, 371 U.S. 471, 484 (1963).  As “warrantless stops and searches are presumptively invalid, the State bears the burden of establishing that any such stop or search is justified by one of the ‘well-delineated exceptions’ to the warrant requirement. ”  State v. Shaw, 213 N.J. 398, 409 (2012) (quoting State v. Frankel, 179 N.J. 586, 598 (2004)).  The law is clear that when a car is stopped due to a purported motor-vehicle violation, “[a]uthority for the seizure . . . ends when tasks tied to the traffic infraction are – or reasonably should have been – completed. \\ 
    \bottomrule
    \end{tabular}
    \label{tab:examples}
\end{table}

\subsection{OPD Resource Library Overview}

The Resource Library is a retrieval system that enables defenders to search across the office’s internal corpus of appellate briefs, directives, and guidelines to locate relevant passages for new cases. The primary goal of the system is to make the collective institutional knowledge of the office, including arguments, citations, and legal reasoning, accessible within seconds. This corpus consists of 2896 briefs spanning the last 25 years, 168 internal documents and 351 public directives. We automatically split these into 140K unique paragraphs using LLM-based semantic segmentation \citep{smith2024evaluating}. 

When a user submits a query, the Library retrieves relevant paragraphs and presents them alongside contextual metadata (such as title and filing date). Each retrieved passage is accompanied by an LLM-generated summary of the legal issue and case facts to help users decide which returned briefs warrant closer inspection. Importantly, generation is only used to summarize existing content and help defenders quickly decide whether a brief might be relevant, not to create new arguments or citations, preserving factual reliability.

Over three evaluation rounds, experienced defenders submitted 194 queries to the tool. For each query, they provided detailed feedback: binary search result annotation (relevant or irrelevant) for up to five retrieved paragraphs, and additional freeform textual feedback. They annotated 85.6\% of all returned paragraphs, from which 66\% were annotated as being relevant for public defense work. Moreover, they provided textual feedback for 55.2\% of the queries, which further contextualize the annotations and search results.

\subsection{The Public Defense Dataset}

Due to confidentiality constraints, the proprietary NJOPD dataset cannot be released. To support reproducible research, we construct the \textbf{Public Defense Dataset} (PD dataset), a resource that mirrors the structure and characteristics of the proprietary dataset with the same search queries, but target paragraphs from publicly available documents. We show dataset examples and annotated paragraphs in Table \ref{tab:examples}.

The PD dataset contains:
\begin{itemize}[nosep]
    \item 170 authentic public defender queries, collected from multiple evaluation rounds with senior public defenders in New Jersey (discarding all queries containing personally identifiable information).
    \item  96,032 unique paragraphs segmented using LLM-based semantic chunking, the same process applied to OPD's internal corpus. They are drawn from a corpus of 856 publicly available documents, including appellate briefs and New Jersey Court/Attorney General directives.
    \item 543 human-annotated relevant paragraphs manually annotated and verified by the author team. All annotators have ample background in legal NLP, and half of the annotators are currently in or have completed law school.    
\end{itemize}

\paragraph{Dataset construction.} To build this public benchmark, we scraped state and AG guidelines, and briefs from cases argued at the appellate level between 2023 and 2025. In total, this results in 856 documents: 351 directives and 505 briefs. We convert all pdfs to text using olmOCR~\citep{olmocrbench}. Afterwards, we apply LLM-based semantic segmentation~\citep{smith2024evaluating} to split the documents into 96,032 unique paragraphs.

We then obtained candidate search results for all queries obtained during the NJOPD evaluation rounds from this corpus (discarding all queries containing personally identifiable information). For each query, we retrieve the following candidates: We collect the 100 most similar paragraphs to the query using the NV-Embed model~\citep{lee2024nv} and the 10 most similar paragraphs based on a keyword search. If we have annotated gold paragraphs from an evaluation round, we also retrieve the 10 most similar paragraphs for each gold paragraph. In total, each query yields 110–160 paragraph candidates: 100 from the NV-Embed, 10 from BM25, and 0–50 from the most similar paragraphs retrieved from annotated results. We narrow down the set of potential candidates using LLMs. First, we use GPT-4o as a judge to filter paragraphs which are potentially irrelevant. We then rerank all remaining paragraphs using a Qwen3-8B reranker \citep{zhang2025qwen3embeddingadvancingtext} fine-tuned on the proprietary NJOPD dataset. 

Next, we manually review up to seven highest scoring paragraphs (if all paragraphs were discarded by the reranker or GPT-4o, we discard the query). For this annotation, we take into account the query, our own legal expertise, the already annotated gold paragraphs from the Resource Library evaluation rounds, and additional feedback collected from those evaluation rounds. The annotation was performed by the author team, all with backgrounds in legal NLP. Additionally, half of the annotators are in or have completed law school. To compute inter-annotator agreement, an independent annotator with law school experience also annotated 100 query/paragraph pairs, using the same annotation guidelines. Annotations between the author team and that annotator result in a Cohen's Kappa of 0.36, indicating fair agreement. Most disagreements (66\% of all disagreements) can be explained by the independent annotator being more lenient, and annotating slightly relevant paragraphs too. The author team was more conservative and often rejected such slightly relevant paragraphs instead.

Lastly, we anonymize party-related personally identifiable information in the dataset. We use the 31B Gemma 4 model \citep{gemma4} to anonymize all paragraphs (prompt shown in Appendix \ref{fig:prompt_anonymization}). This results in 147,044 anonymized entities (1.5 entities per paragraph). To evaluate the anonymization procedure, we manually reviewed 50 paragraphs, 25 where no information was anonymized and 25 paragraphs were at least one entity was anonymized: we did not find any personally identifiable information (PII) in paragraphs where the model did not anonymize. For the anonymized entities, we find that 67\% of anonymizations are correct, while 28\% contain over-anonymization: judge names or organization names insufficient to identify parties. In 5\% of cases the model anonymizes non-relevant legal terms (e.g., case index numbers).

Performance on the two datasets (the proprietary NJOPD and the publicly released PD dataset) is highly correlated: Across eight zero-shot experiments, Recall@5 results in a Spearman R of 0.79 (p = 0.02) and across 20 fine-tuning experiments, the Spearman R is 0.82 (p=9.1e-6). We further show that anonymization barely affects retrieval performance: the Spearman R between the anonymized and non-anonymized version is 1.0 in zero-shot settings (perfect correlation), and 0.99 (p=2.0e-16) for fine-tuning experiments. In Appendix Table \ref{tab:dataset_statistics}, we present dataset statistics from both the PD and the proprietary NJOPD dataset, and show that they are comparable. Combined, these findings make us confident that the PD dataset captures relevant signals about public defender search, and may also be used as a more realistic evaluation benchmark for legal NLP and AI methods.

\section{Taxonomy of Defender Queries}\label{sec:taxonomy}

\begin{figure}[t!]
    \hspace*{0.3cm}
        \centering
    \includegraphics[width=0.7\linewidth]{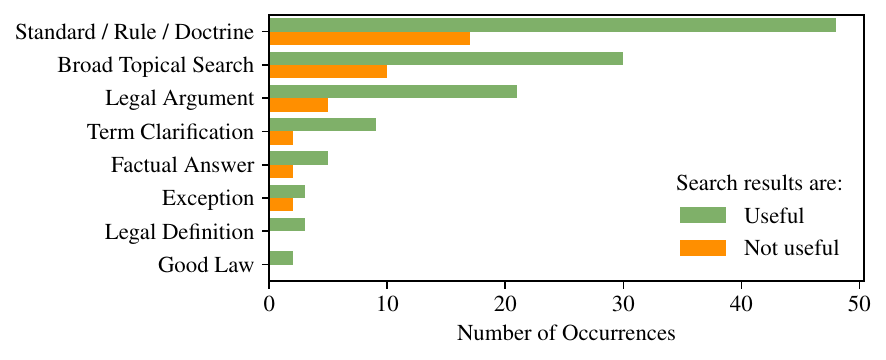}
    \caption{Resource Library performance categorized by \textbf{search intent}. Most queries fall into searching for standards, rules or doctrines, broad topical search, and searching for specific legal arguments.}
    \label{fig:query-objectives}
\end{figure}

To better understand public defense search, we manually annotated all queries for search intent (what a defender was searching for) and algorithmic search strategies (keyword-based, embedding-based, and agentic). We then construct a taxonomy, which includes information about how often queries have been successfully answered by the Resource Library. These can point to common failure cases in realistic legal retrieval. While commercial companies have such data at scale, little to no publicly available information on real-world legal queries can be found. We describe the annotation process in more detail in Appendix Table \ref{tab:query_examples}.

\subsection{Search Intent}

We find substantial variation in public defense search queries, ranging from queries asking for specific cases or rules by title, e.g., \textit{"803(c)(27)"}, to more natural language queries, e.g., \textit{"Find me briefs about inevitable discovery?"} to questions about legality, e.g., \textit{"are tinted windows legal in New Jersey"} to complex queries requiring multihop retrieval and reasoning, such as \textit{"has Counterman v. Colorado been
addressed in a published new jersey opinion?"}. Overall, we identify eight broad categories of search objectives and plot the distributions across them in Figure \ref{fig:query-objectives}. The majority of queries fall into either (1) legal standards, rules or doctrines, (2) search for legal arguments or briefs about certain topics, and (3) less frequently used categories.

\paragraph{Standards, Rules, and Doctrines.} Queries that ask for the legal standard or rules, often for a specific situation, for example \textit{"standard for ordering passenger out of a car"} or \textit{"803(c)(27)"}. These are the most frequent queries, and closely reflect the day-to-day needs of defenders when conducting legal research. They often include keyword-based queries which embedding-based retrieval approaches struggle with.

\paragraph{Topical or Argument-Oriented Searches.} Queries that broadly look for briefs or passages about a certain topic, such as \textit{"find briefs about community caretaking"}, or specifically for legal arguments, for example \textit{"What are arguments against consent searches during illegal car stops?"}.
Having relevant briefs about certain topics can give an overview of the legal landscape and applicable legal arguments. Failure cases include a lack of ability in current models to distinguish nuanced legal contexts: Consider the query \textit{"reverse 404b"}. Reverse 404(b) is when a defense lawyer introduces evidence of another person's past acts to exonerate their client, whereas the standard 404(b) rule is typically used by prosecutors to introduce a defendant's past acts to prove guilt. Embedding models for this query only return results about the standard 404(b) rule.

Other, less frequent categories include term clarifications (e.g., \textit{“difference between reasonable suspicion and probable cause”}), definitions and exceptions (e.g., \textit{“booking exception to Miranda”}), factual or procedural questions (e.g., \textit{“when was NERA amended?”}) and questions about good law (e.g., \textit{"Is statev. pena-flores still good law?"}). Failure cases here can be summarized as the model not sufficiently understanding the query, or the relevant information not being present in the indexed corpus.

\begin{figure}[t!]
    \centering
    \includegraphics[width=0.7\linewidth]{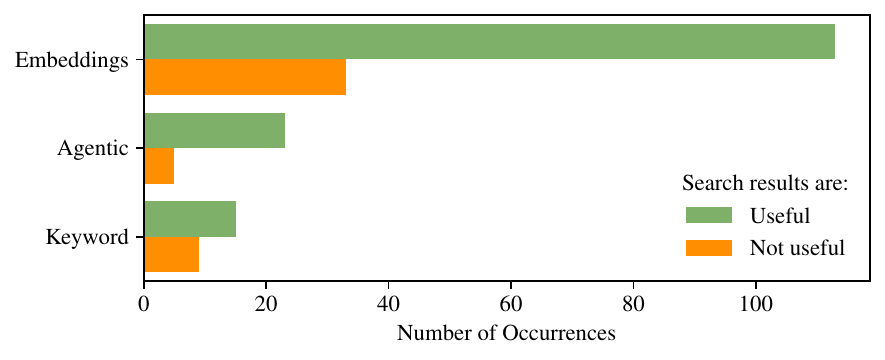}
    \caption{Resource Library performance categorized by \textbf{required search strategy}. Most queries only require embedding-based retrieval. Keyword-based queries are the most challenging for the current system.}
    \label{fig:query-strategy}
\end{figure}

\subsection{Information Sources and Search Strategies.} 
We also annotate the required search strategy for each query. We group search strategies into embedding-based, keyword-based and agentic search (Figure \ref{fig:query-strategy}): 

\begin{itemize}[nosep]
    \item Embedding-based retrieval computes dense vector representations for queries and paragraphs, and the paragraphs closest to a query are returned as results.
    \item Keyword-based retrieval, such as BM25 \citep{bm25s} represent queries and documents as sparse vectors, capturing some statistic about word frequency such as TF-IDF weights. Search results are again paragraphs with the highest similarity to a query.
    \item Agentic search strategies \citep{deng2023mind2webgeneralistagentweb, openai2025operator} refer to methods that broadly involve an agent conducting retrieval, with actions such as query expansion \citep{barexam_qa}, multihop retrieval, and reasoning over multiple retrieval steps. 
\end{itemize} 

Embedding-based retrieval captures most queries, but keyword searches remain common. This is likely due to defenders’ familiarity with boolean-style, commercial legal research systems. Notably, keyword-style queries have the highest rate of unhelpful results (38\%), suggesting that models optimized for natural language retrieval still underperform when users employ traditional legal search syntax. 

One example for an agentic query is \textit{"has Counterman v. Colorado been addressed in a published new jersey opinion"}. To answer, one would first need to retrieve all published opinions referring to Counterman v. Colorado, requiring a search index over all state case law, and associated metadata about publishing status. Then, the agent would need to read all these opinions and decide whether they sufficiently address the main arguments in Counterman v. Colorado, and finally return the answer. Optimally, the agent would provide excerpts from the state opinions addressing the case, along with links to the full opinions. We believe this points to exciting avenues for future work in AI-powered public defense research. 

\section{Empirical Evaluation of Retrieval Models and Rerankers}\label{sec:experiments}
We evaluate two components of the Resource Library pipeline: (1) retrieval models, which encode both queries and paragraphs into embeddings and return a set of candidate results based on semantic similarity, and (2) rerankers, which re-score the top-$k$ retrieved passages using more expressive cross-encoder LLMs.

\begin{table}[t!]
\centering
\caption{Zero-shot Evaluation Metrics}
\label{tab:zeroshot-retrieval}
\footnotesize
\begin{tabular}{l|cc|cc}
\toprule
\textbf{Model} & \multicolumn{2}{c|}{\textbf{NJOPD Dataset}} & \multicolumn{2}{c} {\textbf{PD Dataset}} \\
               & Recall@1 & Recall@5 & Recall@1 & Recall@5  \\
\midrule

all-mpnet-base-v2      & 6.79  & 19.72 & 7.44  & 19.30 \\
E5-base-v2             & 11.11 & 27.44 & 6.27  & 25.23 \\
E5-large-v2            & 11.34 & 29.61 & 8.67  & 27.40 \\
Qwen3-Embedding-0.6B   & 8.69  & 30.93 & 9.28  & 29.35 \\
\hdashline
Qwen3-Embedding-4B     & 10.33 & 36.84 & 11.13 & 34.19 \\
E5-mistral-7b-instruct & 14.48 & 41.97 & 11.11 & 32.61 \\
NV-Embed-v2            & \textbf{15.12} & \textbf{51.85} & 11.48 & 31.27 \\
Qwen3-Embedding-8B     & 13.16 & 40.19 & \textbf{13.37} & \textbf{37.08} \\
\hdashline
fine-tuned E5-large-v2 & 10.72 & 33.71 & 10.40  & 36.26 \\

\bottomrule
\end{tabular}
\par
\vspace{.5em}
\parbox{0.95\linewidth}{
\footnotesize
Evaluation metrics (\%) for various retrieval models with Recall@1 and Recall@5 on the proprietary NJOPD and the PD dataset. Rows sorted by model size: above the dashed line models with less than 1B parameters, below models with more than 1B parameters. Last row shows results of a fine-tuned E5-large model on synthetic data with query expansion, approaching the performance of the much larger Qwen3-Embedding-8B model. See section \ref{sec:experiments} for more details.}

\end{table}

\paragraph{Experimental Setup}
All experiments are conducted on both the proprietary NJOPD dataset, and the PD dataset described in Section~\ref{sec:briefbank}. Each paragraph in the datasets is treated as an independent retrieval unit. Given the size of the datasets, we consider the whole dataset as a test set only and report results on the whole set. For all fine-tuning experiments, we report the mean result of five runs with different seeds (confidence intervals can be found in the Appendix).

\subsection{Retrieval Experiments}
We evaluate retrieval performance of eight pre-trained retrieval models: all-mpnet-base-v2 \citep{reimers-gurevych-2019-sentence}, E5-base-v2 and E5-large-v2 \citep{wang2022text}, Qwen3-Embedding-0.6B, -4B and-8B \citep{zhang2025qwen3embeddingadvancingtext}, E5-mistral-7b-instruct \citep{wang2023improving} and NV-Embed-v2 \citep{lee2024nv}. We use Recall@k (with $k=1,5$) as the metric. This metric is informative for practitioner-facing search systems where defenders typically inspect only the top few results. 

\paragraph{Zero-shot retrieval.} We first evaluate zero-shot performance of eight pre-trained models (Table \ref{tab:zeroshot-retrieval}). We observe that larger models (above 4B parameters) perform better than smaller models (below 1B parameters) on the NJOPD dataset and the PD dataset. The results also confirm that PD is a good approximation of the proprietary NJOPD dataset: performance of the two datasets is correlated, with a Spearman R of 0.79 (p=0.02). If we discard results from the NV-Embed model, we obtain a Spearman R of 0.89 (p=0.007). The NV-Embed model is potentially confounding, as the first version of the Resource Library simply returned the top five paragraphs found by the NV-Embed model. By construction, that model has a recall of 100\% on all these examples, which in turn inflates performance numbers for this model on the proprietary dataset.

\begin{figure}[t!]
    \centering
    \includegraphics[width=0.72\linewidth]{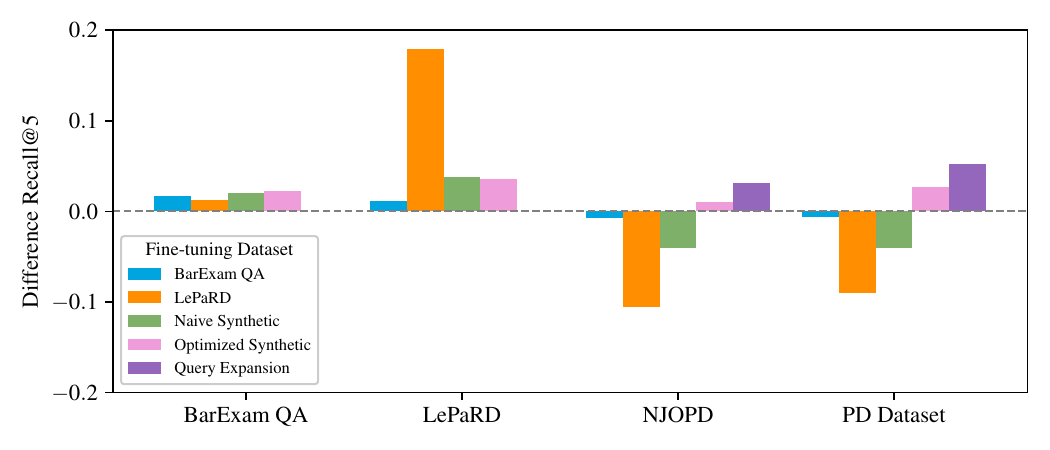}
    \caption{\textbf{Average gains or loss in Recall@5 for four retrieval models.} Colored bars indicate model performance after fine-tuning on different datasets. On the x-axis, we plot resulting model performance on four retrieval datasets (BarExam QA, LePaRD, NJOPD and PD). Changes in recall are relative to a zero-shot baseline of the same model. If fine-tuned on BarExam QA, LePaRD or naive synthetic data, performance on public defense datasets decreases. Fine-tuning on optimized synthetic data or with query expansion strategies increases performance on public defense datasets. Full results are shown in Appendix Table \ref{tab:recalls_training_updated}.}
    \label{fig:training-legal-retrieval-datasets}
\end{figure}

\paragraph{Fine-tuning on existing legal retrieval benchmarks.} Next, we fine-tuned four smaller models (all-mpnet-base-v2, E5-base-v2, E5-large-v2, Qwen3-Embedding-0.6B) on two existing legal retrieval benchmarks: BarExam-QA and LePaRD.
All reported results are the mean of five independent fine-tuning runs with different seeds.  Fine-tuning on these datasets leads to a decrease in Recall@5. In Figure \ref{fig:training-legal-retrieval-datasets}, we show averaged performance gains (or losses). We show exact results for all models in Appendix Table \ref{tab:recalls_training_updated}. To illustrate, the blue bars indicate the effect of further fine-tuning models on the BarExam-QA dataset. If evaluated on the same BarExam-QA dev set, performance increases by 1.7 points in Recall@5, compared to the zero-shot performance of the same models. Training on BarExam-QA also slightly increases performance on LePaRD. However, performance on both the proprietary NJOPD and the released PD dataset decreases.

\paragraph{Fine-tuning on naive synthetic dataset.} Since fine-tuning on existing benchmarks does not lead to performance gain, we experiment with synthetic datasets. We first construct and evaluate a "naive" synthetic dataset generated by a Llama3-70B model \citep{grattafiori2024llama3herdmodels}. To construct the naive synthetic dataset, we generate a corresponding search query for each paragraph in our corpus. To prevent data leakage in this synthetic dataset, we remove all paragraphs which are annotated retrieval targets. We include four (query, paragraph) pairs from the proprietary dataset as few-shot examples to guide generation. The model prompt is shown in Appendix Figure \ref{fig:system_prompt}. Fine-tuning on this dataset increases performance on BarExam-QA and LePARD, indicating that there is some signal about legal similarity in that synthetic dataset. However, training on this dataset substantially decreases performance on the two public defense test sets.

\paragraph{Fine-tuning on optimized synthetic dataset.} Next, we construct an optimized synthetic dataset. We fine-tune a Llama3-70B model using annotated (query, paragraph) pairs obtained in the Resource Library evaluation rounds. Input to the model is an annotated paragraph, output is the query for which the paragraph was retrieved. We use the same system prompt (shown in Appendix Figure \ref{fig:system_prompt}). During this fine-tuning process, the model learns to generate more realistic queries. Next, we filter all (query, paragraph) pairs with a Qwen3-8B reranker fine-tuned on the proprietary NJOPD dataset, and discard all generated examples below a certain threshold. After further inspection of the dataset, we find that paragraphs containing facts, tables of contents, or other procedural content rarely appear in the annotated search results. Thus, we also filter out such paragraphs using a zero-shot Llama3-70B model. Eventually, fine-tuning on the resulting synthetic dataset improves performance on public defense datasets. 

\paragraph{Query Expansion.} Following \citep{barexam_qa}, we also experimented with query expansion strategies. Using Llama3-70B, we expand each query by first applying the IRAC framework (issue, rule, application, conclusion), a well-known method for legal analysis: spot the issue, identify the relevant legal rule, apply the rule to the issue, draw the conclusion. After IRAC, we then derive an expanded search query. Input to the model is the concatenation of the original query, the IRAC analysis and the expanded query. We show the system prompt and one augmented query example in Appendix \ref{tab:query_expansion_prompt}. We expand all public defense queries in the test set, and all curated synthetic queries. Next, we fine-tune models on that set. The E5-large-v2 model approaches the performance of the larger Qwen3-Embedding-8B model in this regime of fine-tuning with expanded queries. In the zero-shot setting, we find mixed results where query expansion increases recall for smaller models, but recall decreases for larger model.

\paragraph{Overall comparison.} Leveraging domain knowledge, we can improve retrieval quality. We show that query expansion eliciting IRAC traces and an optimized synthetic dataset both lead to performance increases. However, training on existing legal benchmarks and naively generated synthetic data decreases quality. We speculate this is caused by a domain shift: these datasets are simply too different from public defense retrieval. We alluded to this phenomenon in Figure \ref{fig:visualization_embeddings} and provide further evidence by showing queries from all datasets in Appendix Table \ref{tab:dataset-examples}. We show exact results (instead of average performance) for all models in Appendix Table \ref{tab:recalls_training_updated}.

\paragraph{Robustness of the PD Dataset} Results between the proprietary NJOPD dataset and the released PD dataset are highly correlated. In zero-shot settings, the resulting Spearman R is 0.79 (p=0.02) for Recall@5, for fine-tuning experiments, the Spearman R consists of 0.82 (p=9.1e-6) across 20 fine-tuning experiments (four models and five different training datasets). We also report retrieval results on a non-anonymized version of the PD dataset, and report results in Appendix Table \ref{tab:results_non_anonymized}. The resulting Spearman R of the anonymized and non-anonymized version is 1.0 in the zero-shot setting (perfect correlation), the correlation after 20 fine-tuning experiments results in a Spearman R of 0.99 (p=2.0e-16). We believe these results together confirm the validity of the benchmark, although it has been annotated by the author team and undergone anonymization.

Since NJOPD documents span 25 years, and our PD dataset only contains briefs from 2023--2025, we also verify that retrieval performance is robust across different time periods of the NJOPD dataset. We report Spearman R between the PD dataset and NJOPD subsets stratified by year in Appendix Table \ref{tab:opd-by-year}. Correlations between the datasets are not driven by old NJOPD documents, but also generalize to more recent retrieval targets.

In all fine-tuning experiments, we train models five times with different seeds. In the main text, we consistently report the mean results of these five runs. In Appendix Table \ref{tab:recalls_training_CI}, we show the 95\% confidence intervals. In almost all experiments, differences to zero-shot results of the same base model are significant. The confidence intervals of the fine-tuning experiments are moderate, with an average of 0.70 percentage points for Recall@5 for the NJOPD dataset, and an average of 0.83 percentage points for the PD dataset. These are comparable to the confidence intervals of the BarExam QA dataset (0.68 percentage points).

Given the practitioner-facing nature of the PD dataset, we report Recall@5 as the official metric (the Resource Library by default returns five search results). Recall@5 is highly correlated with other standard information retrieval metrics, including Normalized Discounted Cumulative Gain (NDCG@5), Mean Reciprocal Rank (MRR@10), and Mean Average Precision (MAP@100). The lowest of these correlations with the reported Recall@5 results is between MRR@10 and Recall@5, with a Spearman R of 0.97 (p=2.6e-12). We show detailed results for all metrics in Appendix Table \ref{tab:retrieval_metrics_with_spearman}.

\paragraph{Domain Adaptation.} We additionally experiment with legal domain adaptation, another method to add domain knowledge to models. We take a ModernBERT-large checkpoint further pre-trained on 30B tokens of US case law opinions using the masked language modeling objective from \citep{stammbach2026legaldomainadaptationmodern}. Fine-tuning this domain-adapted checkpoint on the optimized synthetic dataset achieves a Recall@5 of 23.4, outperforming a vanilla ModernBERT-large checkpoint fine-tuned on the same dataset (Recall@5=18.7, -4.7 percentage points). We show exact results in Appendix Table \ref{tab:modernbert_results}.

\subsection{Reranker Experiments}

\begin{figure}[t!]
    \centering
    \includegraphics[width=0.8\linewidth]{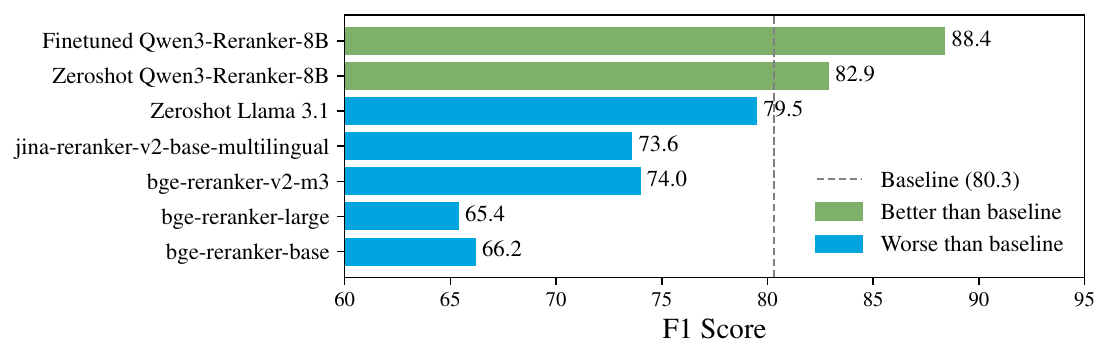}
    \caption{\textbf{F1 scores of different reranker models.} A majority baseline achieves an F1 of 80.3 (most annotated paragraphs are relevant, so a majority baseline achieves a high F1 score). Most off-the-shelf rerankers perform worse than this majority baseline.}
    \label{fig:rerankers}
\end{figure}

We evaluate six open-source rerankers on the NJOPD dataset. We additionally fine-tune the most performant reranker on the proprietary dataset, showing that this further improves performance. We use an 80-20 split to separate queries into a training and test set (stratified by queries, to make sure that no training queries appear in the test set). We then calculate the F1 score for all models. We show the main results in Figure \ref{fig:rerankers}.

We compare all results to a simple majority baseline. The baseline assumes a simple heuristic which treats each search result as a good result. We plot the F1 score of this majority baseline as a horizontal dashed line (at 80.3\% F1). We observe that most existing rerankers underperform the simple majority baseline, with the only exception being the recently released Qwen3-8B reranker \citep{zhang2025qwen3embeddingadvancingtext}, which slightly surpasses this baseline. In contrast, fine-tuning on domain-specific data leads to significant improvements. Detailed metrics (precision, recall, accuracy, and F1) for all models are provided in Appendix Table \ref{tab:rerankers}.

\section{Discussion}\label{sec:discussion}

\subsection{Difference between Existing Benchmarks and Public Defense Retrieval} 

We find a distribution mismatch between existing legal retrieval datasets and our public defender search dataset, where training on legal benchmarks \citep{barexam_qa, mahari-etal-2024-lepard} results in lower performance. Similarly, \citet{gu2025illusionreadinessstresstesting} note that outdated medical domain benchmarks are inadequate to assess current AI, and \citet{ott2022mapping} emphasize that most existing benchmarks lack real-world utility. Especially for public defense work, there seem to be no suitable benchmarks to advance AI and NLP methods. Moreover, most legal retrieval datasets do not contain manually verified retrieval targets, and none contain real-world queries (Appendix Table \ref{tab:legal_retrieval_datasets}). While the field of NLP and AI for public good and access to justice is growing \citep{karamolegkou2025nlpsocialgoodsurvey, mahari-etal-2023-law, mahari-etal-2024-lepard}, such efforts are limited by the availability of realistic data.

To make progress on this front, we provide several starting points to stimulate further research on public defense retrieval. We release the PD dataset containing realistic queries drafted by experienced public defenders, and manually verified corresponding paragraphs relevant to those queries. Performance on the PD dataset is correlated with a proprietary dataset created by public defenders. Next, we construct a taxonomy about search objectives of defenders and what search strategies can be employed to answer these queries. Similar to datasets like WildChat \citep{zhao2024wildchat1mchatgptinteraction}, which make real-world ChatGPT conversations accessible for research, we hope our dataset and taxonomy can inform future work in more realistic legal retrieval settings.

\subsection{Collaborations between Academia and Legal Institutions}

This work illustrates a collaboration between academic researchers and a public agency. We believe such partnerships can be mutually beneficial: Public agencies get to be involved in reflecting on their existing workflows and AI needs, and gain exposure to the opportunities, trade-offs, risks and barriers of AI applications.

For researchers, such collaborations allow work on AI tasks and use cases that are more closely aligned with real institutional needs and constraints. These are often absent in existing datasets and AI benchmarks, especially in the legal domain (See Appendix Table \ref{tab:legal_retrieval_datasets}). Through frequent meetings and discussions, knowledge transfer benefits both sides: agencies clarify their AI needs, and researchers gain insight into tacit constraints that can inform better AI methods.

We believe such collaborations represent a promising research direction to work on more realistic AI applications with practical utility \citep{ott2022mapping}. As NLP and AI techniques mature and increasingly promise real-world impact, progress is often limited by the lack of realistic datasets and evaluation settings. We view our collaboration with NJOPD, and the accompanying data artifacts presented in this paper as one step toward addressing this gap.

\subsection{Future Work To Improve Public Defense Retrieval}

\citet{barexam_qa} find substantial gains leveraging query expansion and legal reasoning to retrieve relevant statutes for bar exam questions. We confirm this for public defense retrieval, and believe there's value in exploring such efforts in more detail. Moreover, we believe agentic search \citep{deng2023mind2webgeneralistagentweb}, which combines multiple retrieval steps into a single agentic workflow. During this process, the agent can explore e.g., query expansion, legal reasoning, multihop retrieval, and reranking, until a suitable search result is retrieved.

Second, the current state-of-the-art embedding models, i.e., E5-mistral \citep{wang2023improving}, Qwen3-Embedding \citep{zhang2025qwen3embeddingadvancingtext} and NV-Embed-v2 \citep{lee2024nv} already perform substantially better than smaller, older models built on top of BERT or RoBERTa models, as shown in Table \ref{tab:zeroshot-retrieval}. We believe further such advances, and especially advances in dedicated legal retrieval models, may further increase performance on public defense retrieval.

Synthetic data likely can be leveraged to further improve model performance in legal retrieval. However, we note two challenges: The first is that synthetic data must be carefully curated, as we have shown in Section \ref{sec:experiments}. In Section \ref{sec:AI_for_public_defenders}, we described how NJOPD has experimented with a RAG tool, but ultimately rejected the idea: models often were imprecise in addressing specific legal questions. It appears that current models lack the ability to handle nuanced legal contexts \citep[see also][]{cheong2025aiaugmentaccessjustice}, which will affect the generated data. This remains an open research problem.

\subsection{Related Work}

Our work contributes to a growing academic field on how to use legal NLP and AI in collaboration with public agencies. Related work includes AI assistance for automatically detecting and redacting racial covenants \citep{surani2025aiscalinglegalreform}, automatically clearing records at scale \citep{codeforamerica2020clearmyrecord} or AI assistance for eviction defense \citep{stanford2025empowering_legalaid}. Similar to our work, these projects also identified a suitable use case for AI assistance, and in collaboration with public agencies developed specialized methods to accomplish the goal.

Moreover, we release a dataset for legal retrieval, and contribute to the academic literature on legal retrieval in the United States \citep{coliee_2022_summary, mahari-etal-2024-lepard, hou-etal-2025-clerc, barexam_qa}. In contrast to these works, we focus on retrieval for public defenders, where the goal is to retrieve relevant paragraphs from legal briefs. Following the recommendations made in \citet{ott2022mapping}, we have put an emphasis on real-world utility while creating this benchmark.

We do acknowledge the potential of NLP and AI methods for other types of public defense work. \citet{cheong2025aiaugmentaccessjustice} outline a research agenda for how AI can assist public defenders, and foremost identify use cases around making sense of large volumes of data in evidence investigation. Similar to our work, public defenders in \citep{cheong2025aiaugmentaccessjustice} report similar challenges of using AI for legal research and writing, however this can change rapidly as technology advances.

\section{Conclusion}\label{sec:conclusion}

In this paper, we discuss legal retrieval for public defenders. In collaboration with the New Jersey Office of the Public Defender, we identify retrieval over internal briefs as a suitable use case to assist public defense work, and developed the OPD Resource Library. The development of this tool allowed us to gather realistic public defense queries, from which we manually construct and release the PD retrieval dataset.

Query expansion, carefully curated synthetic datasets and legal domain adaptation increase retrieval performance, while in-domain fine-tuning increases reranking accuracy. However, training on existing academic legal retrieval datasets lowers performance, indicating a distribution shift between these benchmarks and the more realistic PD retrieval task. Our results suggest that progress in legal retrieval for public defense may be constrained less by model scale than by domain mismatch and lack of available datasets.

\section{Acknowledgements}

We thank all our partners at the New Jersey Office of the Public Defenders, most importantly Jennifer Perez and Alison Perrone without whom this work would not have been possible. Moreover, we would like to thank Walker Gosrich from NJ Innovation Authority and John Seith from the NJ Office of Information Technology for their technical assistance, and Joe Krakora for his insights and guidance throughout the project.

\section{Broader Impact Statement}

This work introduces public defense retrieval as an NLP task and releases an associated benchmark. We have built the OPD Resource Library, an AI-powered retrieval tool to surface relevant appellate briefs to support defenders' day-to-day legal research, while taking into account high stakes and low error tolerance of public defense work. The tool is deliberately designed to not be a AI decision-making system, but AI assistance mimicking existing office practices. The tool serves as an interface to the office's collective institutional knowledge and mirrors existing word-of-mouth practices.

The Library avoids using generative AI for substantive legal tasks. Exploratory evaluation revealed unacceptable failure rates in RAG-based approaches (hallucinations, failure to handle nuanced legal contexts), consistent with qualitative findings defenders report in \citet{cheong2025aiaugmentaccessjustice}. LLM-generated summaries are included in the tool only to help defenders triage results. We implement multiple additional safeguards to encourage responsible usage: (1) We specify in the Library instructions that defenders shall not rely on AI-generated summaries for details, but for triaging whether returned documents are relevant. (2) The summaries are not visible by default, but have to be expanded. (3) If expanded, a user first sees a disclaimer (in bold and large font-size) that the summary is AI-generated and might be incomplete, factually incorrect or contain hallucinations. (4) We added CSS safeguards to prevent that these summaries can be copy-pasted.

The released PD dataset enables future research in more realistic legal retrieval settings. Performance on the public dataset is strongly correlated with a proprietary defender-annotated dataset, suggesting that it provides a valid signal for benchmarking. We hope it also serves as a starting point for similar deployments at other defender offices. However, we acknowledge that the benchmark stems from a single office in one U.S. state. Generalization to other jurisdictions or office contexts should thus always be verified. Moreover, the benchmark might be biased (1) towards retrieval targets specific to New Jersey, (2) specific annotation idiosyncrasies of the annotating defenders and the author team, and (3) we relied on automatic filtering using an ensemble of GPT-4o and a fine-tuned Qwen3-8B reranker to further reduce the number of paragraphs we manually reviewed. These limitations need to be considered in future usage of this benchmark.

Finally, any deployment of AI tools in public defense must carefully consider confidentiality. Using proprietary APIs risks falling outside attorney-client privilege and may be subject to mandatory disclosure \citep{cheong2024not, privilege}. The Resource Library was deployed as a closed system within existing secure NJ State infrastructure. Thus, the tool is only accessible to NJOPD attorneys, and relies on open-source models. We recommend other projects developing access to justice AI tools to take into account constraints around sensitive data, privilege and confidentiality, and to adopt similar safeguards.

\bibliography{main}

@String{Computing = "Computing" }

@String{Computer = "{IEEE} Computer" }

@String{Chelsea = "Chelsea" }

@String{Springer = "Springer-Verlag" }

@misc{bommasani2022opportunitiesrisksfoundationmodels,
      title={On the Opportunities and Risks of Foundation Models}, 
      author={Rishi Bommasani and Drew A. Hudson and Ehsan Adeli and Russ Altman and Simran Arora and Sydney von Arx and Michael S. Bernstein and Jeannette Bohg and Antoine Bosselut and Emma Brunskill and Erik Brynjolfsson and Shyamal Buch and Dallas Card and Rodrigo Castellon and Niladri Chatterji and Annie Chen and Kathleen Creel and Jared Quincy Davis and Dora Demszky and Chris Donahue and Moussa Doumbouya and Esin Durmus and Stefano Ermon and John Etchemendy and Kawin Ethayarajh and Li Fei-Fei and Chelsea Finn and Trevor Gale and Lauren Gillespie and Karan Goel and Noah Goodman and Shelby Grossman and Neel Guha and Tatsunori Hashimoto and Peter Henderson and John Hewitt and Daniel E. Ho and Jenny Hong and Kyle Hsu and Jing Huang and Thomas Icard and Saahil Jain and Dan Jurafsky and Pratyusha Kalluri and Siddharth Karamcheti and Geoff Keeling and Fereshte Khani and Omar Khattab and Pang Wei Koh and Mark Krass and Ranjay Krishna and Rohith Kuditipudi and Ananya Kumar and Faisal Ladhak and Mina Lee and Tony Lee and Jure Leskovec and Isabelle Levent and Xiang Lisa Li and Xuechen Li and Tengyu Ma and Ali Malik and Christopher D. Manning and Suvir Mirchandani and Eric Mitchell and Zanele Munyikwa and Suraj Nair and Avanika Narayan and Deepak Narayanan and Ben Newman and Allen Nie and Juan Carlos Niebles and Hamed Nilforoshan and Julian Nyarko and Giray Ogut and Laurel Orr and Isabel Papadimitriou and Joon Sung Park and Chris Piech and Eva Portelance and Christopher Potts and Aditi Raghunathan and Rob Reich and Hongyu Ren and Frieda Rong and Yusuf Roohani and Camilo Ruiz and Jack Ryan and Christopher Ré and Dorsa Sadigh and Shiori Sagawa and Keshav Santhanam and Andy Shih and Krishnan Srinivasan and Alex Tamkin and Rohan Taori and Armin W. Thomas and Florian Tramèr and Rose E. Wang and William Wang and Bohan Wu and Jiajun Wu and Yuhuai Wu and Sang Michael Xie and Michihiro Yasunaga and Jiaxuan You and Matei Zaharia and Michael Zhang and Tianyi Zhang and Xikun Zhang and Yuhui Zhang and Lucia Zheng and Kaitlyn Zhou and Percy Liang},
      year={2022},
      eprint={2108.07258},
      archivePrefix={arXiv},
      primaryClass={cs.LG},
      url={https://arxiv.org/abs/2108.07258}, 
}

@article{lee2024nv,
  title={NV-Embed: Improved Techniques for Training LLMs as Generalist Embedding Models},
  author={Lee, Chankyu and Roy, Rajarshi and Xu, Mengyao and Raiman, Jonathan and Shoeybi, Mohammad and Catanzaro, Bryan and Ping, Wei},
  journal={arXiv preprint arXiv:2405.17428},
  year={2024}
}

@misc{grattafiori2024llama3herdmodels,
      title={The Llama 3 Herd of Models}, 
      author={Aaron Grattafiori and Abhimanyu Dubey and Abhinav Jauhri and Abhinav Pandey and Abhishek Kadian and Ahmad Al-Dahle and Aiesha Letman and Akhil Mathur and Alan Schelten and Alex Vaughan and Amy Yang and Zhiyu Ma et al.},
      year={2024},
      eprint={2407.21783},
      archivePrefix={arXiv},
      primaryClass={cs.AI},
      url={https://arxiv.org/abs/2407.21783}, 
}

@techreport{smith2024evaluating,
  title = {Evaluating Chunking Strategies for Retrieval},
  author = {Smith, Brandon and Troynikov, Anton},
  year = {2024},
  month = {July},
  institution = {Chroma},
  url = {https://research.trychroma.com/evaluating-chunking},
}

@misc{justia_gideon_v._wainwright,
author = {{Gideon v. Wainwright}},
  title = {{372 U.S. 335}},
  url = {https://supreme.justia.com/cases/federal/us/372/335/},
  urldate = {2024-07-14},
  year = {1963},
}

@techreport{Pace2023,
  title={National Public Defense Workload Study},
  author={Nicholas M. Pace and Malia N. Brink and Cynthia G. Lee and Stephen F. Hanlon},
  year={2023},
  institution={RAND Corporation}
}

@misc{bm25s,
      title={BM25S: Orders of magnitude faster lexical search via eager sparse scoring}, 
      author={Xing Han Lù},
      year={2024},
      eprint={2407.03618},
      archivePrefix={arXiv},
      primaryClass={cs.IR},
      url={https://arxiv.org/abs/2407.03618}, 
}

@misc{zhang2025qwen3embeddingadvancingtext,
      title={Qwen3 Embedding: Advancing Text Embedding and Reranking Through Foundation Models}, 
      author={Yanzhao Zhang and Mingxin Li and Dingkun Long and Xin Zhang and Huan Lin and Baosong Yang and Pengjun Xie and An Yang and Dayiheng Liu and Junyang Lin and Fei Huang and Jingren Zhou},
      year={2025},
      eprint={2506.05176},
      archivePrefix={arXiv},
      primaryClass={cs.CL},
      url={https://arxiv.org/abs/2506.05176}, 
}

@inproceedings{mahari-etal-2024-lepard,
    title = "{L}e{P}a{RD}: A Large-Scale Dataset of Judicial Citations to Precedent",
    author = "Mahari, Robert  and
      Stammbach, Dominik  and
      Ash, Elliott  and
      Pentland, Alex",
    editor = "Ku, Lun-Wei  and
      Martins, Andre  and
      Srikumar, Vivek",
    booktitle = "Proceedings of the 62nd Annual Meeting of the Association for Computational Linguistics (Volume 1: Long Papers)",
    month = aug,
    year = "2024",
    address = "Bangkok, Thailand",
    publisher = "Association for Computational Linguistics",
    url = "https://aclanthology.org/2024.acl-long.532/",
    doi = "10.18653/v1/2024.acl-long.532",
    pages = "9863--9877"
}

@article{vandermaaten08a,
  author  = {Laurens van der Maaten and Geoffrey Hinton},
  title   = {Visualizing Data using t-SNE},
  journal = {Journal of Machine Learning Research},
  year    = {2008},
  volume  = {9},
  number  = {86},
  pages   = {2579--2605},
  url     = {http://jmlr.org/papers/v9/vandermaaten08a.html}
}

@inproceedings{chen-etal-2025-air,
    title = "{AIR}-Bench: Automated Heterogeneous Information Retrieval Benchmark",
    author = "Chen, Jianlyu  and
      Wang, Nan  and
      Li, Chaofan  and
      Wang, Bo  and
      Xiao, Shitao  and
      Xiao, Han  and
      Liao, Hao  and
      Lian, Defu  and
      Liu, Zheng",
    editor = "Che, Wanxiang  and
      Nabende, Joyce  and
      Shutova, Ekaterina  and
      Pilehvar, Mohammad Taher",
    booktitle = "Proceedings of the 63rd Annual Meeting of the Association for Computational Linguistics (Volume 1: Long Papers)",
    month = jul,
    year = "2025",
    address = "Vienna, Austria",
    publisher = "Association for Computational Linguistics",
    url = "https://aclanthology.org/2025.acl-long.982/",
    pages = "19991--20022",
    ISBN = "979-8-89176-251-0"
}

@misc{deng2023mind2webgeneralistagentweb,
      title={Mind2Web: Towards a Generalist Agent for the Web}, 
      author={Xiang Deng and Yu Gu and Boyuan Zheng and Shijie Chen and Samuel Stevens and Boshi Wang and Huan Sun and Yu Su},
      year={2023},
      eprint={2306.06070},
      archivePrefix={arXiv},
      primaryClass={cs.CL},
      url={https://arxiv.org/abs/2306.06070}, 
}

@techreport{openai2025operator,
  title        = {Operator System Card},
  author       = {{OpenAI}},
  institution  = {OpenAI},
  year         = {2025},
  month        = jan,
  day          = {23},
  type         = {System Card},
  url          = {https://cdn.openai.com/operator_system_card.pdf}
}

@article{wang2022text,
  title={Text Embeddings by Weakly-Supervised Contrastive Pre-training},
  author={Wang, Liang and Yang, Nan and Huang, Xiaolong and Jiao, Binxing and Yang, Linjun and Jiang, Daxin and Majumder, Rangan and Wei, Furu},
  journal={arXiv preprint arXiv:2212.03533},
  year={2022}
}

@article{wang2023improving,
  title={Improving Text Embeddings with Large Language Models},
  author={Wang, Liang and Yang, Nan and Huang, Xiaolong and Yang, Linjun and Majumder, Rangan and Wei, Furu},
  journal={arXiv preprint arXiv:2401.00368},
  year={2023}
}

@inproceedings{reimers-gurevych-2019-sentence,
    title = "Sentence-{BERT}: Sentence Embeddings using {S}iamese {BERT}-Networks",
    author = "Reimers, Nils  and
      Gurevych, Iryna",
    editor = "Inui, Kentaro  and
      Jiang, Jing  and
      Ng, Vincent  and
      Wan, Xiaojun",
    booktitle = "Proceedings of the 2019 Conference on Empirical Methods in Natural Language Processing and the 9th International Joint Conference on Natural Language Processing (EMNLP-IJCNLP)",
    month = nov,
    year = "2019",
    address = "Hong Kong, China",
    publisher = "Association for Computational Linguistics",
    url = "https://aclanthology.org/D19-1410/",
    doi = "10.18653/v1/D19-1410",
    pages = "3982--3992"
}

@inproceedings{hou-etal-2025-clerc,
    title = "{CLERC}: A Dataset for {U}. {S}. Legal Case Retrieval and Retrieval-Augmented Analysis Generation",
    author = "Hou, Abe Bohan  and
      Weller, Orion  and
      Qin, Guanghui  and
      Yang, Eugene  and
      Lawrie, Dawn  and
      Holzenberger, Nils  and
      Blair-Stanek, Andrew  and
      Van Durme, Benjamin",
    editor = "Chiruzzo, Luis  and
      Ritter, Alan  and
      Wang, Lu",
    booktitle = "Findings of the Association for Computational Linguistics: NAACL 2025",
    month = apr,
    year = "2025",
    address = "Albuquerque, New Mexico",
    publisher = "Association for Computational Linguistics",
    url = "https://aclanthology.org/2025.findings-naacl.441/",
    doi = "10.18653/v1/2025.findings-naacl.441",
    pages = "7898--7913",
    ISBN = "979-8-89176-195-7"
}

@misc{gao2021scalingdeepcontrastivelearning,
      title={Scaling Deep Contrastive Learning Batch Size under Memory Limited Setup}, 
      author={Luyu Gao and Yunyi Zhang and Jiawei Han and Jamie Callan},
      year={2021},
      eprint={2101.06983},
      archivePrefix={arXiv},
      primaryClass={cs.LG},
      url={https://arxiv.org/abs/2101.06983}, 
}

@article{edwards2019affirmation,
  author       = {Barry C. Edwards},
  title        = {Why Appeals Courts Rarely Reverse Lower Courts: An Experimental Study to Explore Affirmation Bias},
  journal      = {Emory Law Journal Online},
  volume       = {68},
  pages        = {1035--1073},
  year         = {2019},
  url          = {https://scholarlycommons.law.emory.edu/elj-online/7},
}

@misc{stanford2025empowering_legalaid,
  author       = {{Stanford Law School Legal Design Lab}},
  title        = {Empowering Legal Aid: Developing AI Co‑Pilots for Eviction Defense and Reentry Debt Mitigation},
  year         = {2025},
  howpublished = {Press release / in‑brief article by Stanford Law School},
  note         = {Stanford Law School, “Empowering Legal Aid” initiative description; accessed August 3, 2025},
  url          = {https://law.stanford.edu/press/empowering-legal-aid/}
}

@misc{surani2025aiscalinglegalreform,
      title={AI for Scaling Legal Reform: Mapping and Redacting Racial Covenants in Santa Clara County}, 
      author={Faiz Surani and Mirac Suzgun and Vyoma Raman and Christopher D. Manning and Peter Henderson and Daniel E. Ho},
      year={2025},
      eprint={2503.03888},
      archivePrefix={arXiv},
      primaryClass={cs.CL},
      url={https://arxiv.org/abs/2503.03888}, 
}

@misc{codeforamerica2020clearmyrecord,
  author       = {{Code for America}},
  title        = {Record Clearance at Scale: How Clear My Record Helped Reduce or Dismiss 144,000 Convictions in {California}},
  year         = {2020},
  note         = {Accessed: 2025-08-03}
}

@inproceedings{promptagator,
  author       = {Zhuyun Dai and
                  Vincent Y. Zhao and
                  Ji Ma and
                  Yi Luan and
                  Jianmo Ni and
                  Jing Lu and
                  Anton Bakalov and
                  Kelvin Guu and
                  Keith B. Hall and
                  Ming{-}Wei Chang},
  title        = {Promptagator: Few-shot Dense Retrieval From 8 Examples},
  booktitle    = {The Eleventh International Conference on Learning Representations,
                  {ICLR} 2023, Kigali, Rwanda, May 1-5, 2023},
  publisher    = {OpenReview.net},
  year         = {2023},
  url          = {https://openreview.net/forum?id=gmL46YMpu2J},
  bibsource    = {dblp computer science bibliography, https://dblp.org}
}

@misc{karamolegkou2025nlpsocialgoodsurvey,
      title={NLP for Social Good: A Survey of Challenges, Opportunities, and Responsible Deployment}, 
      author={Antonia Karamolegkou and Angana Borah and Eunjung Cho and Sagnik Ray Choudhury and Martina Galletti and Rajarshi Ghosh and Pranav Gupta and Oana Ignat and Priyanka Kargupta and Neema Kotonya and Hemank Lamba and Sun-Joo Lee and Arushi Mangla and Ishani Mondal and Deniz Nazarova and Poli Nemkova and Dina Pisarevskaya and Naquee Rizwan and Nazanin Sabri and Dominik Stammbach and Anna Steinberg and David Tomás and Steven R Wilson and Bowen Yi and Jessica H Zhu and Arkaitz Zubiaga and Anders Søgaard and Alexander Fraser and Zhijing Jin and Rada Mihalcea and Joel R. Tetreault and Daryna Dementieva},
      year={2025},
      eprint={2505.22327},
      archivePrefix={arXiv},
      primaryClass={cs.CL},
      url={https://arxiv.org/abs/2505.22327}, 
}

@misc{gu2025illusionreadinessstresstesting,
      title={The Illusion of Readiness: Stress Testing Large Frontier Models on Multimodal Medical Benchmarks}, 
      author={Yu Gu and Jingjing Fu and Xiaodong Liu and Jeya Maria Jose Valanarasu and Noel Codella and Reuben Tan and Qianchu Liu and Ying Jin and Sheng Zhang and Jinyu Wang and Rui Wang and Lei Song and Guanghui Qin and Naoto Usuyama and Cliff Wong and Cheng Hao and Hohin Lee and Praneeth Sanapathi and Sarah Hilado and Bian Jiang and Javier Alvarez-Valle and Mu Wei and Jianfeng Gao and Eric Horvitz and Matt Lungren and Hoifung Poon and Paul Vozila},
      year={2025},
      eprint={2509.18234},
      archivePrefix={arXiv},
      primaryClass={cs.AI},
      url={https://arxiv.org/abs/2509.18234}, 
}

@misc{stammbach2026legaldomainadaptationmodern,
      title={Legal Domain Adaptation of Modern BERT Models}, 
      author={Dominik Stammbach and Peter Henderson},
      year={2026},
      eprint={2606.28538},
      archivePrefix={arXiv},
      primaryClass={cs.CL},
      url={https://arxiv.org/abs/2606.28538}, 
}

@inproceedings{coliee_2022_summary,
author = {Kim, Mi-Young and Rabelo, Juliano and Goebel, Randy and Yoshioka, Masaharu and Kano, Yoshinobu and Satoh, Ken},
title = {COLIEE 2022 Summary: Methods For Legal Document Retrieval And Entailment},
year = {2023},
isbn = {978-3-031-29167-8},
publisher = {Springer-Verlag},
address = {Berlin, Heidelberg},
url = {https://doi.org/10.1007/978-3-031-29168-5_4},
doi = {10.1007/978-3-031-29168-5_4},
booktitle = {New Frontiers in Artificial Intelligence: JSAI-IsAI 2022 Workshop, JURISIN 2022, and JSAI 2022 International Session, Kyoto, Japan, June 12–17, 2022, Revised Selected Papers},
pages = {51–67},
numpages = {17},
location = {Kyoto, Japan}
}

@inproceedings{barexam_qa,
author = {Zheng, Lucia and Guha, Neel and Arifov, Javokhir and Zhang, Sarah and Skreta, Michal and Manning, Christopher D. and Henderson, Peter and Ho, Daniel E.},
title = {A Reasoning-Focused Legal Retrieval Benchmark},
year = {2025},
isbn = {9798400714214},
publisher = {Association for Computing Machinery},
address = {New York, NY, USA},
url = {https://doi.org/10.1145/3709025.3712219},
doi = {10.1145/3709025.3712219},
booktitle = {Proceedings of the 2025 Symposium on Computer Science and Law},
pages = {169–193},
numpages = {25},
location = {Munich, Germany},
series = {CSLAW '25}
}

@article{sklearn,
author = {Pedregosa, Fabian and Varoquaux, Ga\"{e}l and Gramfort, Alexandre and Michel, Vincent and Thirion, Bertrand and Grisel, Olivier and Blondel, Mathieu and Prettenhofer, Peter and Weiss, Ron and Dubourg, Vincent and Vanderplas, Jake and Passos, Alexandre and Cournapeau, David and Brucher, Matthieu and Perrot, Matthieu and Duchesnay, \'{E}douard},
title = {Scikit-learn: Machine Learning in Python},
year = {2011},
issue_date = {2/1/2011},
publisher = {JMLR.org},
volume = {12},
number = {null},
issn = {1532-4435},
journal = {J. Mach. Learn. Res.},
month = nov,
pages = {2825–2830},
numpages = {6}
}

@misc{cheong2025aiaugmentaccessjustice,
      title={How Can AI Augment Access to Justice? Public Defenders' Perspectives on AI Adoption}, 
      author={Inyoung Cheong and Patty Liu and Dominik Stammbach and Peter Henderson},
      year={2025},
      eprint={2510.22933},
      archivePrefix={arXiv},
      primaryClass={cs.CY},
      url={https://arxiv.org/abs/2510.22933}, 
}

@misc{dominguezolmedo2025lawmapowerspecializationlegal,
      title={Lawma: The Power of Specialization for Legal Annotation}, 
      author={Ricardo Dominguez-Olmedo and Vedant Nanda and Rediet Abebe and Stefan Bechtold and Christoph Engel and Jens Frankenreiter and Krishna Gummadi and Moritz Hardt and Michael Livermore},
      year={2025},
      eprint={2407.16615},
      archivePrefix={arXiv},
      primaryClass={cs.CL},
      url={https://arxiv.org/abs/2407.16615}, 
}

@article{SchwarczChoi2023AITools,
  author    = {Schwarcz, Daniel and Choi, Jonathan H.},
  title     = {AI Tools for Lawyers: A Practical Guide},
  journal   = {Minnesota Law Review Headnotes},
  volume    = {108},
  pages     = {1},
  year      = {2023},
  note      = {Minnesota Legal Studies Research Paper},
  month     = mar,
  url       = {https://ssrn.com/abstract=4404017},
  doi       = {10.2139/ssrn.4404017}
}

@inproceedings{manor-li-2019-plain,
  address = {Minneapolis, Minnesota},
  author = {Manor, Laura  and
Li, Junyi Jessy},
  booktitle = {Proceedings of the Natural Legal Language Processing Workshop 2019},
  month = jun,
  pages = {1--11},
  publisher = {Association for Computational Linguistics},
  title = {Plain {E}nglish Summarization of Contracts},
  url = {https://www.aclweb.org/anthology/W19-2201},
  year = {2019},
}

@misc{butler2025massivelegalembeddingbenchmark,
      title={The Massive Legal Embedding Benchmark (MLEB)}, 
      author={Umar Butler and Abdur-Rahman Butler and Adrian Lucas Malec},
      year={2025},
      eprint={2510.19365},
      archivePrefix={arXiv},
      primaryClass={cs.CL},
      url={https://arxiv.org/abs/2510.19365}, 
}

@inproceedings{muennighoff-etal-2023-mteb,
    title = "{MTEB}: Massive Text Embedding Benchmark",
    author = "Muennighoff, Niklas  and
      Tazi, Nouamane  and
      Magne, Loic  and
      Reimers, Nils",
    editor = "Vlachos, Andreas  and
      Augenstein, Isabelle",
    booktitle = "Proceedings of the 17th Conference of the European Chapter of the Association for Computational Linguistics",
    month = may,
    year = "2023",
    address = "Dubrovnik, Croatia",
    publisher = "Association for Computational Linguistics",
    url = "https://aclanthology.org/2023.eacl-main.148/",
    doi = "10.18653/v1/2023.eacl-main.148",
    pages = "2014--2037"
}

@misc{zhang2025llmstrulyunderstandprecedent,
      title={Do LLMs Truly Understand When a Precedent Is Overruled?}, 
      author={Li Zhang and Jaromir Savelka and Kevin Ashley},
      year={2025},
      eprint={2510.20941},
      archivePrefix={arXiv},
      primaryClass={cs.CL},
      url={https://arxiv.org/abs/2510.20941}, 
}

@inproceedings{10.1145/3462757.3466066,
author = {Huang, Zihan and Low, Charles and Teng, Mengqiu and Zhang, Hongyi and Ho, Daniel E. and Krass, Mark S. and Grabmair, Matthias},
title = {Context-aware legal citation recommendation using deep learning},
year = {2021},
isbn = {9781450385268},
publisher = {Association for Computing Machinery},
address = {New York, NY, USA},
url = {https://doi.org/10.1145/3462757.3466066},
doi = {10.1145/3462757.3466066},
booktitle = {Proceedings of the Eighteenth International Conference on Artificial Intelligence and Law},
pages = {79–88},
numpages = {10},
location = {S\~{a}o Paulo, Brazil},
series = {ICAIL '21}
}

@article{dadgostari2021modeling,
  title={Modeling law search as prediction},
  author={Dadgostari, Faraz and Guim, Mauricio and Beling, Peter A and Livermore, Michael A and Rockmore, Daniel N},
  journal={Artificial Intelligence and Law},
  volume={29},
  number={1},
  pages={3--34},
  year={2021},
  publisher={Springer}
}

@article{10.1098/rsta.2023.0254,
    author = {Katz, Daniel Martin and Bommarito, Michael James and Gao, Shang and Arredondo, Pablo},
    title = {GPT-4 passes the bar exam},
    journal = {Philosophical Transactions of the Royal Society A: Mathematical, Physical and Engineering Sciences},
    volume = {382},
    number = {2270},
    pages = {20230254},
    year = {2024},
    month = {02},
    issn = {1364-503X},
    doi = {10.1098/rsta.2023.0254},
    url = {https://doi.org/10.1098/rsta.2023.0254},
    eprint = {https://royalsocietypublishing.org/rsta/article-pdf/doi/10.1098/rsta.2023.0254/1328474/rsta.2023.0254.pdf},
}

@inproceedings{10.5555/3666122.3668037,
author = {Guha, Neel and Nyarko, Julian and Ho, Daniel E. and R\'{e}, Christopher and Chilton, Adam and Narayana, Aditya and Chohlas-Wood, Alex and Peters, Austin and Waldon, Brandon and Rockmore, Daniel N. and Zambrano, Diego and Talisman, Dmitry and Hoque, Enam and Surani, Faiz and Fagan, Frank and Sarfaty, Galit and Dickinson, Gregory M. and Porat, Haggai and Hegland, Jason and Wu, Jessica and Nudell, Joe and Niklaus, Joel and Nay, John and Choi, Jonathan H. and Tobia, Kevin and Hagan, Margaret and Ma, Megan and Livermore, Michael and Rasumov-Rahe, Nikon and Holzenberger, Nils and Kolt, Noam and Henderson, Peter and Rehaag, Sean and Goel, Sharad and Gao, Shang and Williams, Spencer and Gandhi, Sunny and Zur, Tom and Iyer, Varun and Li, Zehua},
title = {LEGALBENCH: a collaboratively built benchmark for measuring legal reasoning in large language models},
year = {2023},
publisher = {Curran Associates Inc.},
address = {Red Hook, NY, USA},
booktitle = {Proceedings of the 37th International Conference on Neural Information Processing Systems},
articleno = {1915},
numpages = {157},
location = {New Orleans, LA, USA},
series = {NIPS '23}
}

@misc{olmocrbench,
      title={{olmOCR: Unlocking Trillions of Tokens in PDFs with Vision Language Models}},
      author={Jake Poznanski and Jon Borchardt and Jason Dunkelberger and Regan Huff and Daniel Lin and Aman Rangapur and Christopher Wilhelm and Kyle Lo and Luca Soldaini},
      year={2025},
      eprint={2502.18443},
      archivePrefix={arXiv},
      primaryClass={cs.CL},
      url={https://arxiv.org/abs/2502.18443},
}

@misc{zhao2024wildchat1mchatgptinteraction,
      title={WildChat: 1M ChatGPT Interaction Logs in the Wild}, 
      author={Wenting Zhao and Xiang Ren and Jack Hessel and Claire Cardie and Yejin Choi and Yuntian Deng},
      year={2024},
      eprint={2405.01470},
      archivePrefix={arXiv},
      primaryClass={cs.CL},
      url={https://arxiv.org/abs/2405.01470}, 
}

@inproceedings{mahari-etal-2023-law,
    title = "The Law and {NLP}: Bridging Disciplinary Disconnects",
    author = "Mahari, Robert  and
      Stammbach, Dominik  and
      Ash, Elliott  and
      Pentland, Alex",
    editor = "Bouamor, Houda  and
      Pino, Juan  and
      Bali, Kalika",
    booktitle = "Findings of the Association for Computational Linguistics: EMNLP 2023",
    month = dec,
    year = "2023",
    address = "Singapore",
    publisher = "Association for Computational Linguistics",
    url = "https://aclanthology.org/2023.findings-emnlp.224/",
    doi = "10.18653/v1/2023.findings-emnlp.224",
    pages = "3445--3454"
}

@article{privilege,
    author = {Adam Paine and Robert M. Travisano},
    journal = {Epstein Becker Green}, 
    title = {Discovery Pitfalls in the Age of AI},
    howpublished = {\url{https://techcrunch.com/2025/07/25/sam-altman-warns-theres-no-legal-confidentiality-when-using-chatgpt-as-a-therapist/}},
    year = {2025}, 
    month = {September}}

@inproceedings{cheong2024not,
  title={(A) I am not a lawyer, but...: engaging legal experts towards responsible LLM policies for legal advice},
  author={Cheong, Inyoung and Xia, King and Feng, KJ Kevin and Chen, Quan Ze and Zhang, Amy X},
  booktitle={Proceedings of the 2024 ACM Conference on Fairness, Accountability, and Transparency},
  pages={2454--2469},
  year={2024}
}

@misc{gemma4,
  title        = {Gemma 4},
  author       = {{Google DeepMind}},
  year         = {2026},
  howpublished = {\url{https://deepmind.google/models/gemma/gemma-4/}},
  note         = {Accessed: 2026-04-30}
}

@article{ott2022mapping,
  title={Mapping global dynamics of benchmark creation and saturation in artificial intelligence},
  author={Ott, Simon and Barbosa-Silva, Adriano and Blagec, Kathrin and Brauner, Jan and Samwald, Matthias},
  journal={Nature Communications},
  volume={13},
  number={1},
  pages={6793},
  year={2022},
  publisher={Nature Publishing Group UK London}
}
\bibliographystyle{tmlr}

\newpage
\appendix
\section{Appendix}

\begin{table}[h!]
\centering
\caption{Comparison of Queries and Paragraphs in Legal Retrieval Datasets}
\label{tab:legal_retrieval_datasets}
\scriptsize
\begin{tabular}{lccc}
\toprule
\textbf{Dataset} & \textbf{Domain} & \textbf{Real queries} & \textbf{Manually verified targets}  \\
\midrule
LePaRD \citep{mahari-etal-2024-lepard} & US case law & \xmark & \xmark \\
CLERC \citep{hou-etal-2025-clerc} & US case law & \xmark & \xmark \\
Law Search as Prediction \citep{dadgostari2021modeling} & US case law & \xmark & \xmark \\
AirBench \citep{chen-etal-2025-air} & Pile of Law & \xmark & \xmark \\ 
BVA citation prediction \citep{10.1145/3462757.3466066} & Board of Veterans' Appeals & \xmark & \xmark \\
Contract Summarization \citep{manor-li-2019-plain} & Contracts & \xmark & \xmark \\
COLIEE \citep{coliee_2022_summary} & US case law / statutes & \xmark & \cmark \\
BarExam-QA \citep{barexam_qa} & US case law / bar exams & \xmark & \cmark \\ \hdashline
PD Dataset (ours) & Appellate briefs & \cmark & \cmark \\

\bottomrule
\end{tabular}
\par
\vspace{.5em}
\parbox{0.95\linewidth}{
\footnotesize
Comparison of queries and paragraphs in legal retrieval datasets. This list includes all retrieval datasets  used in LegalBench \citep{10.5555/3666122.3668037}, all U.S. legal retrieval datasets in MTE-Bench \citep{muennighoff-etal-2023-mteb} and all U.S. datasets in MLEB \citep{butler2025massivelegalembeddingbenchmark}.
}
\end{table}

\begin{table}[h!]
\centering
\caption{Dataset Statistics}
\label{tab:dataset_statistics}

\footnotesize
\begin{tabular}{lcc}
\toprule
\textbf{Statistic} & \textbf{Proprietary Dataset} & \textbf{Released Dataset} \\
\midrule
Number of queries       & 194 & 170   \\
Average gold paragraphs per query & 2.9 & 3.2 \\
Average query length (words)    & 8.9 & 9.3  \\
Average paragraph length (words) & 133.9 & 155.0  \\
Type-Token Ratio (TTR)  & 0.15 & 0.13   \\

\bottomrule
\end{tabular}
\end{table}

\begin{table}[h!]
    \centering
    \caption{Examples of Queries and Annotations for Taxonomy Construction}    
    \label{tab:query_examples}
    
    \scriptsize
    \begin{tabular}{p{0.45\textwidth} p{0.22\textwidth} p{0.22\textwidth}} 
    \toprule
    \textbf{Query} & \textbf{Objective} & \textbf{Search Strategy} \\ \midrule
    Standard for ordering passenger out of a car & standard & embeddings \\
    803(c)(27) & rule & keyword \\
    inevitable discovery & doctrine & keywords / embeddings \\
    find briefs about community caretaking & topical search & keywords / embeddings \\
    What are arguments against consent searches during illegal car stops? & legal argument & embeddings\\
    Difference between reasonable suspicion and probable cause & term clarification & embeddings \\
    when was NERA  amended to enumerate the offenses subject to its provisions? & Factual Answer & agentic \\
    has Counterman v. Colorado been addressed in a published new jersey opinion? & Factual Answer & agentic \\
    booking exception to miranda & exception & embeddings  \\
    what is the definition of probable cause? & definition & embeddings \\
    Is state v. pena-flores still good law? & good law & embeddings / agentic \\ \bottomrule
    \end{tabular}

    \vspace{0.5em}
\par
\vspace{.5em}
\parbox{0.95\linewidth}{
\footnotesize
Examples of Queries and Annotations for the purposes of deriving a taxonomy. We manually annotate all queries in our dataset. While annotating, we consider the following information: The query itself, relevant search results annotated by experienced public defenders for a query which help us understand what they were searching for, and additional freeform textual feedback. Examples in this table were used as annotation guidelines.
}

\end{table}

\begin{table}[h]
\centering
\caption{Spearman R Correlations between NJOPD Subsets (Stratified by Release Date) and the PD Dataset}
\label{tab:opd-by-year}
\footnotesize
\begin{tabular}{lccc}
\toprule
\textbf{NJOPD Period} & \textbf{Queries} & \textbf{Retrieval Targets} & \textbf{Spearman R (p-value)} \\
\midrule
All years (overall) & 194  & 563 & 0.79 (0.02) \\
1998--2019          & 114  & 357 & 1.00 ($<$0.001) \\
2020--2025          & 70   & 137 & 0.88 (0.004) \\
2023--2025          & 46   &  66 & 0.61 (0.108) \\
\bottomrule
\end{tabular}
\parbox{0.95\linewidth}{
\footnotesize
The NJOPD corpus spans 25 years and its annotated retrieval targets are predominantly from older briefs; the PD dataset covers documents from 2023--2025. Strong correlations across all time periods indicate that retrieval performance on annotated paragraphs of older NJOPD documents generalizes to newer ones. The marginally significant Spearman R of 0.61 between the PD dataset and the NJOPD subset is likely driven by the small dataset size of the corresponding NJOPD subset. We exclude documents where the year information is missing from this analysis.}
\end{table}

\begin{table}
\centering
\caption{Recall@5 Results after Fine-tuning Models on Various Legal Retrieval Datasets}
\label{tab:recalls_training_updated}

\footnotesize
\begin{tabular}{llcccc}
\toprule
\textbf{Train Dataset} & \textbf{Model} & \textbf{BarExam QA} & \textbf{LePaRD} & \textbf{Internal NJOPD} & \textbf{PD Dataset} \\
\midrule

\multirow{4}{*}{BarExam QA}
  & all-mpnet-base-v2    & 4.52 (+2.90)   & 15.75 (+1.44)  & 21.80 (+2.07)  & 19.67 (+0.37) \\
  & E5-base-v2           & 3.23 (-1.61)   & 16.82 (+0.47)  & 28.06 (+0.62)  & 25.46 (+0.23) \\
  & E5-large-v2          & 3.23 (+1.61)   & 18.64 (+1.11)  & 32.03 (+2.43)  & 28.11 (+0.71) \\
  & Qwen3-0.6B           & \textbf{11.94} (+3.87) & 20.51 (+1.41) & 22.66 (-8.26) & 25.37 (-3.98) \\
\hdashline
\multirow{4}{*}{LePaRD}
  & all-mpnet-base-v2    & 3.39 (+1.77)   & 32.03 (+17.71) & 18.17 (-1.56) & 17.35 (-1.96) \\
  & E5-base-v2           & 4.84 (+0.00)   & 28.82 (+12.47) & 9.59 (-17.85) & 7.46 (-17.77) \\
  & E5-large-v2          & 5.97 (+4.35)   & 36.52 (+18.99) & 20.01 (-9.59) & 19.41 (-7.99) \\
  & Qwen3-0.6B           & 7.10 (-0.97)   & \textbf{41.70} (+22.60) & 18.08 (-12.84) & 21.04 (-8.32) \\
\hdashline
\multirow{4}{*}{Naive Synthetic}
  & all-mpnet-base-v2    & 5.97 (+4.35)   & 19.81 (+5.49)  & 23.39 (+3.67) & 22.42 (+3.12) \\
  & E5-base-v2           & 3.39 (-1.45)   & 18.62 (+2.28)  & 20.37 (-7.07) & 17.57 (-7.66) \\
  & E5-large-v2          & 8.71 (+7.10)   & 21.14 (+3.60)  & 25.35 (-4.25) & 23.76 (-3.64) \\
  & Qwen3-0.6B           & 6.61 (-1.45)   & 22.90 (+3.80)  & 22.46 (-8.47) & 21.45 (-7.91) \\
\hdashline
\multirow{4}{*}{Optimized Synthetic}
  & all-mpnet-base-v2    & 7.90 (+6.29)   & 19.68 (+5.37)  & 27.28 (+7.56) &
  24.53 (+5.23) \\
  & E5-base-v2           & 3.23 (-1.61)   & 18.23 (+1.88)  & 27.99 (+0.54) & 24.42 (-0.80) \\
  & E5-large-v2          & 7.26 (+5.65)   & 21.28 (+3.74)  & 29.84 (+0.24) & 31.09 (+3.68) \\
  & Qwen3-0.6B           & 6.77 (-1.29)   & 22.31 (+3.21)  & 26.54 (-4.38) & 31.99 (+2.64) \\
\hdashline
\multirow{4}{*}{Query Expansion}
  & all-mpnet-base-v2    & n/a & n/a & 31.89 (+12.17) & 26.48 (+7.18) \\
  & E5-base-v2           & n/a & n/a & 31.65 (+4.21)  & 27.82 (+2.6) \\
  & E5-large-v2          & n/a & n/a & \textbf{33.71} (+4.10) & \textbf{36.26} (+8.86) \\
  & Qwen3-0.6B           & n/a & n/a & 27.03 (-3.90) & 35.32 (+5.97) \\
  
\bottomrule
\end{tabular}
\par
\vspace{.5em}
\parbox{0.95\linewidth}{
\footnotesize
Recall@5 results obtained via training models on different legal retrieval datasets. In brackets: difference to zero-shot version of the same model. All models trained with \texttt{sentence-transformers} \citep{reimers-gurevych-2019-sentence}, using a learning rate of 2e-5, a batch size of 128, and the CachedMultipleNegativesRankingLoss \citep{gao2021scalingdeepcontrastivelearning}. Best performance by benchmark in bold (Qwen3-0.6B trained on BarExam QA for BarExam, Qwen3-0.6B trained on LePaRD for LePaRD, E5-large-v2 trained on query expansions for both public defender search datasets.)
}
\end{table}

\begin{table}
\centering
\caption{Recall@5 Results with Confidence Intervals}
\label{tab:recalls_training_CI}

\footnotesize
\begin{tabular}{llcccc}
\toprule
\textbf{Train Dataset} & \textbf{Model} & \textbf{BarExam QA} & \textbf{LePaRD} & \textbf{Internal NJOPD} & \textbf{PD Dataset} \\
\midrule

\multirow{4}{*}{BarExam QA}
  & all-mpnet-base-v2    & 4.52 ($\pm{0.9}$)**   & 15.75 ($\pm{0.04}$)**  & 21.80 ($\pm{0.25}$)**  & 19.67 ($\pm{0.32}$)** \\
  & E5-base-v2           & 3.23 ($\pm{0.0}$)**   & 16.82 ($\pm{0.03}$)**  & 28.06 ($\pm{0.32}$)**  & 25.46 ($\pm{0.54}$)** \\
  & E5-large-v2          & 3.23 ($\pm{0.0}$)**   & 18.64 ($\pm{0.03}$)**  & 32.03 ($\pm{0.64}$)**  & 28.11 ($\pm{0.21}$)** \\
  & Qwen3-0.6B           & \textbf{11.94} ($\pm{0.84}$)** & 20.51 ($\pm{0.07}$)** & 22.66 ($\pm{0.60}$)** & 25.37 ($\pm{0.73}$)** \\
\hdashline
\multirow{4}{*}{LePaRD}
& all-mpnet-base-v2    & 3.39 ($\pm{0.45}$)**  & 32.03 ($\pm{0.02}$)** & 18.17 ($\pm{0.33}$)** & 17.35 ($\pm{0.46}$)** \\
  & E5-base-v2           & 4.84 ($\pm{0.0}$)** & 28.82 ($\pm{0.02}$)** & 9.59 ($\pm{0.61}$)**  & 7.46 ($\pm{0.47}$)** \\
  & E5-large-v2          & 5.97 ($\pm{1.82}$)**  & 36.52 ($\pm{0.08}$)** & 20.01 ($\pm{0.67}$)** & 19.41 ($\pm{1.29}$)** \\
  & Qwen3-0.6B           & 7.10 ($\pm{1.49}$)    & \textbf{41.70} ($\pm{0.17}$)** & 18.08 ($\pm{1.29}$)** & 21.04 ($\pm{1.27}$)** \\

\hdashline
\multirow{4}{*}{Naive Synthetic}
& all-mpnet-base-v2    & 5.97 ($\pm{0.55}$)** & 19.81 ($\pm{0.03}$)** & 23.39 ($\pm{0.15}$)** & 22.42 ($\pm{0.33}$)** \\
  & E5-base-v2           & 3.39 ($\pm{0.45}$)** & 18.62 ($\pm{0.05}$)** & 20.37 ($\pm{0.48}$)** & 17.57 ($\pm{0.45}$)** \\
  & E5-large-v2          & 8.71 ($\pm{1.10}$)** & 21.14 ($\pm{0.12}$)** & 25.35 ($\pm{0.98}$)** & 23.76 ($\pm{2.25}$)** \\
  & Qwen3-0.6B           & 6.61 ($\pm{1.31}$)** & 22.90 ($\pm{0.11}$)** & 22.46 ($\pm{1.63}$)** & 21.45 ($\pm{1.67}$)** \\

\hdashline
\multirow{4}{*}{Optimized Synthetic}
 & all-mpnet-base-v2    & 7.90 ($\pm{0.45}$)** & 19.68 ($\pm{0.04}$)** & 27.28 ($\pm{0.41}$)** & 24.53 ($\pm{0.63}$)** \\
  & E5-base-v2           & 3.23 ($\pm{0.0}$)**  & 18.23 ($\pm{0.08}$)** & 27.99 ($\pm{0.36}$)** & 24.42 ($\pm{0.99}$) \\
  & E5-large-v2          & 7.26 ($\pm{0.71}$)** & 21.28 ($\pm{0.08}$)** & 29.84 ($\pm{1.37}$)   & 31.09 ($\pm{0.55}$)** \\
  & Qwen3-0.6B           & 6.77 ($\pm{0.90}$)** & 22.31 ($\pm{0.05}$)** & 26.54 ($\pm{0.62}$)** & 31.99 ($\pm{0.49}$)** \\

\hdashline
\multirow{4}{*}{Query Expansion}
 & all-mpnet-base-v2    & n/a & n/a & 31.89 ($\pm{0.46}$)** & 26.48 ($\pm{0.74}$)** \\
  & E5-base-v2           & n/a & n/a & 31.65 ($\pm{0.71}$)** & 27.82 ($\pm{0.95}$)** \\
  & E5-large-v2          & n/a & n/a & \textbf{33.71} ($\pm{1.09}$)** & \textbf{36.26} ($\pm{0.53}$)** \\
  & Qwen3-0.6B           & n/a & n/a & 27.03 ($\pm{1.06}$) ** & 35.32 ($\pm{1.85}$)** \\
  
\bottomrule
\end{tabular}
\par
\vspace{.5em}
\parbox{0.95\linewidth}{
\footnotesize
Recall@5 results obtained via training models on different legal retrieval datasets. In brackets: 95\% confidence intervals across 5 independent fine-tuning runs with different seeds. All models trained with \texttt{sentence-transformers} \citep{reimers-gurevych-2019-sentence}, using a learning rate of 2e-5, a batch size of 128, and the CachedMultipleNegativesRankingLoss \citep{gao2021scalingdeepcontrastivelearning}. Best performance by benchmark in bold (Qwen3-0.6B trained on BarExam QA for BarExam, Qwen3-0.6B trained on LePaRD for LePaRD, E5-large-v2 trained on query expansions for both public defender search datasets.)}

\end{table}

\begin{table*}
\centering
\caption{Different IR Evaluation Metrics}
\label{tab:retrieval_metrics_with_spearman}
\tiny
\begin{tabular}{llcccc|cccc}
\toprule
\textbf{Train Dataset} & \textbf{Model}
& \multicolumn{4}{c}{\textbf{Internal NJOPD}}
& \multicolumn{4}{c}{\textbf{PD Dataset}} \\
\cline{3-10}
& & R@5 & NDCG@5 & MRR@10 & MAP@100
  & R@5 & NDCG@5 & MRR@10 & MAP@100 \\
\midrule

\multirow{4}{*}{BarExam QA}
& all-mpnet-base-v2 & 21.80 & 21.62 & 36.46 & 21.00 & 19.76 & 19.34 & 31.16 & 17.96 \\
& E5-base-v2        & 28.06 & 29.30 & 49.30 & 27.44 & 25.41 & 23.25 & 35.48 & 21.40 \\
& E5-large-v2       & 32.03 & 33.22 & 51.29 & 32.09 & 28.16 & 26.56 & 40.38 & 24.58 \\
& Qwen3-0.6B        & 22.66 & 21.37 & 35.56 & 21.53 & 25.51 & 23.83 & 36.58 & 22.43 \\
\hdashline

\multirow{4}{*}{LePaRD}
& all-mpnet-base-v2 & 18.17 & 18.80 & 33.96 & 18.24 & 17.28 & 16.21 & 26.73 & 15.52 \\
& E5-base-v2        & 9.59  & 9.52  & 18.37 & 8.98  & 7.54  & 7.37  & 13.22 & 7.03 \\
& E5-large-v2       & 20.01 & 19.36 & 34.30 & 17.95 & 19.58 & 19.06 & 31.53 & 17.21 \\
& Qwen3-0.6B        & 18.08 & 17.45 & 30.94 & 16.46 & 21.41 & 19.60 & 29.43 & 18.60 \\
\hdashline

\multirow{4}{*}{Naive Synthetic}
& all-mpnet-base-v2 & 23.39 & 23.60 & 39.14 & 23.62 & 22.42 & 20.32 & 29.32 & 19.35 \\
& E5-base-v2        & 20.37 & 20.78 & 34.20 & 19.82 & 17.53 & 16.11 & 25.58 & 16.13 \\
& E5-large-v2       & 25.35 & 24.98 & 39.87 & 24.22 & 23.78 & 22.05 & 34.75 & 21.52 \\
& Qwen3-0.6B        & 22.46 & 21.58 & 36.02 & 21.47 & 21.50 & 20.61 & 33.06 & 20.22 \\
\hdashline

\multirow{4}{*}{Optimized Synthetic}
& all-mpnet-base-v2 & 27.28 & 27.71 & 46.07 & 27.74 & 24.63 & 22.88 & 34.89 & 22.81 \\
& E5-base-v2        & 27.99 & 28.71 & 46.46 & 27.49 & 24.44 & 23.53 & 36.64 & 22.93 \\
& E5-large-v2       & 29.84 & 29.85 & 46.70 & 30.35 & 30.80 & 28.37 & 42.26 & 27.92 \\
& Qwen3-0.6B        & 26.54 & 26.67 & 42.89 & 26.71 & 31.88 & 29.25 & 42.45 & 28.53 \\
\hdashline

\multirow{4}{*}{Query Expansion}
& all-mpnet-base-v2 & 31.89 & 34.72 & 55.58 & 34.52 & 26.54 & 24.99 & 38.15 & 25.51 \\
& E5-base-v2        & 31.65 & 31.86 & 48.58 & 30.85 & 27.94 & 26.08 & 38.85 & 25.56 \\
& E5-large-v2       & 33.71 & 33.90 & 51.29 & 34.45 & 36.0 & 32.28 & 46.23 & 30.48 \\
& Qwen3-0.6B        & 27.03 & 27.36 & 44.21 & 27.78 & 35.10 & 32.11 & 45.62 & 31.19 \\
\midrule

\multicolumn{2}{l}{\textbf{Spearman R}}
& -- & 0.989 & 0.983 & 0.979 
& -- & 0.992 & 0.968 & 0.974 \\
\bottomrule
\end{tabular}
\parbox{0.95\linewidth}{
\footnotesize
We report different Information Retrieval (IR) evaluation metrics of various fine-tuned models on the public defense retrieval dataset. In the last row, we report Spearman R between the reported metric and the official Recall@5 metric, which we have been reporting throughout the main text of the paper. All correlations among different metrics are almost perfect and highly significant.}
\end{table*}

\begin{table}
\centering
\caption{Domain Adaptation Results}
\label{tab:modernbert_results}

\footnotesize
\begin{tabular}{llcccc}
\toprule
\textbf{Train Dataset} & \textbf{Model} & \textbf{BarExam QA} & \textbf{LePaRD} & \textbf{NJOPD} & \textbf{PD Dataset} \\
\midrule
\multirow{2}{*}{BarExam QA}
  & ModernBERT-large & 1.61 ($\pm$0.00) & 7.30 ($\pm$0.06) & 1.02 ($\pm$0.18) & 0.26 ($\pm$0.00) \\
  & Legal-ModernBERT & \textbf{2.42 ($\pm$0.00)} & \textbf{8.77 ($\pm$0.04)} & 1.18 ($\pm$0.00) & \textbf{1.78 ($\pm$0.23)} \\
\hdashline
\multirow{2}{*}{LePaRD}
  & ModernBERT-large & 3.55 ($\pm$0.90) & 32.16 ($\pm$0.12) & 6.85 ($\pm$0.40) & 5.25 ($\pm$0.60) \\
  & Legal-ModernBERT & 3.23 ($\pm$0.00) & \textbf{32.41 ($\pm$0.12)} & \textbf{9.46 ($\pm$0.90)} & \textbf{9.89 ($\pm$0.17)} \\
\hdashline
\multirow{2}{*}{Naive Synthetic}
  & ModernBERT-large & 3.71 ($\pm$0.55) & 17.30 ($\pm$0.11) & 10.88 ($\pm$0.27) & 11.75 ($\pm$1.11) \\
  & Legal-ModernBERT & 4.03 ($\pm$0.00) & \textbf{17.80 ($\pm$0.05)} & \textbf{12.48 ($\pm$0.51)} & \textbf{16.0 ($\pm$0.83)} \\
\hdashline
\multirow{2}{*}{Optimized Synthetic}
  & ModernBERT-large & 2.42 ($\pm$0.00) & 17.01 ($\pm$0.08) & 20.34 ($\pm$0.74) & 18.74 ($\pm$0.53) \\
  & Legal-ModernBERT & \textbf{4.19 ($\pm$0.45)} & \textbf{17.42 ($\pm$0.05)} & 20.82 ($\pm$1.11) & \textbf{23.39 ($\pm$0.83)} \\
\bottomrule
\end{tabular}
\par
\vspace{.5em}
\parbox{0.95\linewidth}{
\footnotesize
Recall@5 with 95\% confidence intervals (mean $\pm$ CI over five seeds). Bold indicates statistically significant differences between Legal-ModernBERT and ModernBERT-large, based on non-overlapping confidence intervals. All models trained with \texttt{sentence-transformers} using a learning rate of 2e-5, batch size 128, and CachedMultipleNegativesRankingLoss.
}
\end{table}

\begin{table}[]
    \centering
    \caption{Comparison Retrieval Performance (Recall@5) on Anonymized and Non-Anonymized PD Dataset}
    \label{tab:results_non_anonymized}
    \scriptsize

    \begin{tabular}{l c c c }
    \toprule
    \textbf{Experiment} & \textbf{Anonymized} & \textbf{Non-anonymized} & $\Delta$ \textbf{(Difference)} \\ \midrule
    all-mpnet-base-v2 & 19.30 & 19.55 & $-$ 0.25 \\
    E5-base-v2 & 25.23 & 25.06 & $+$ 0.17 \\
    E5-large-v2 & 27.40 & 28.25 & $-$ 0.85 \\
    Qwen3-Embedding-0.6B & 29.35 & 29.83 & $-$0.48 \\
    Qwen3-Embedding-4B & 34.19 & 34.10 & $+$0.09 \\
    E5-mistral-7b-instruct  & 32.61 & 33.63 & $-$1.02 \\
    NV-Embed-v2 & 31.27 & 31.93 & $-$0.66 \\
    Qwen3-Embedding-8B & 37.08 & 37.37 & $-$0.29 \\ \bottomrule
    \end{tabular}
\parbox{0.95\linewidth}{
\footnotesize
Spearman R of 1.0 (perfect correlation) for retrieval performance between anonymized and non-anonymized version of the PD dataset (zero-shot settings). Spearman R across 20 fine-tuning experiments (See Table \ref{tab:recalls_training_updated}) is 0.99 (p=2.0e-16).}    
\end{table}

\begin{table}
\centering
\caption{Detailed Reranker Results}
\label{tab:rerankers}
\footnotesize
\begin{tabular}{l|cccc|cccc}
\toprule
\textbf{Model} & \multicolumn{4}{c|}{\textbf{All (\%)}} & \multicolumn{4}{c}{\textbf{Heldout Test (\%)}} \\
               & Pr   & Rc   & F1   & Acc  & Pr   & Rc   & F1   & Acc \\
\midrule
majority baseline & 66.8 & 1.00 & 80.1 & 66.8 & 67.1 & 1.00 & 80.3 & 67.1 \\
bge-reranker-base & 77.4 & 59.8 & 67.5 & 62.1 & 76.5 & 58.4 & 66.2 & 62.4 \\
bge-reranker-large & 77.5 & 56.3 & 65.2 & 60.5 & 78.1 & 56.2 & 65.4 & 62.4 \\
bge-reranker-v2-m3 & 75.7 & 73.6 & 74.6 & 67.1 & 76.2 & 71.9 & 74.0 & 68.1 \\
jina-reranker-v2-base-multilingual & 76.4 & 65.1 & 70.3 & 63.8 & 81.1 & 67.4 & 73.6 & 69.5 \\
Zeroshot Llama 3.1 & 77.9 & 81.7 & 79.7 & 72.7 & 80.5 & 78.7 & 79.5 & 74.5 \\
Zeroshot Qwen3-Reranker-8B & 73.7 & 98.7 & 84.4 & 76.0 & 71.9 & 97.8 & 82.9 & 74.5 \\
\hdashline
Finetuned-Qwen3-Reranker-8B & n/a & n/a & n/a & n/a & 87.0 & 89.9 & 88.4 & 85.1 \\

\bottomrule
\end{tabular}
\par
\vspace{.5em}
\parbox{0.95\linewidth}{
\footnotesize
Evaluation metrics (\%) for various reranker models. For zero-shot rerankers, we show performance on all datapoints and a heldout test set, for fine-tuned models, we only show performance for 20\% randomly held-out datapoints.}

\end{table}

\begin{table}[h!]
\centering
\caption{Examples Queries from Different Datasets}
\label{tab:dataset-examples}
\tiny
\renewcommand{\arraystretch}{1.3} 
\begin{tabular}{l|p{12.5cm}}
\toprule
\textbf{Dataset} & \textbf{Queries} \\
\midrule

\multirow{3}{*}{BarExam QA} 
& Paul, the Plaintiff in a personal injury action, called Wes as a witness to testify that Dan's car, in which Paul had been riding. ran a red light. Wes, however, testified that Dan's car did not run the light. Paul then called Vic to testify that Dan's car did run the light. The trial judge should rule that Vic's testimony is \\
& Paul, the Plaintiff in a personal injury action, called Wes as a witness to testify that Dan's car, in which Paul had been riding. ran a red light. Wes, however, testified that Dan's car did not run the light. On cross-examination of Vic, Dan's attorney asked if Vic was drunk at the time he witnessed the accident. and Vic responded, "No I have never in my life been drunk." Dan's attorney then sought to prove by Yank that Vic was drunk on New Year's Eve two years before the accident. The trial judge should rule that Yank's testimony is
 \\
& Paul, the Plaintiff in a personal injury action, called Wes as a witness to testify that Dan's car, in which Paul had been riding. ran a red light. Wes, however, testified that Dan's car did not run the light. Dan called Zemo as a witness and asked him if he knew Vic's reputation for veracity in the community where Vic resided. The trial judge should rule that this question is \\
\hline

\multirow{3}{*}{LePaRD} 
& Hill, 131 F.3d at 1062 (quoting Mathis, 963 F.2d at 408). In this case, because Smith’s and Cook’s escape indictments are devoid of detail, and Thomas’ indictments were never proffered to the court, the parties agree that we should look no further than the statutory language. See Taylor, 495 U.S. at 600, 110 S.Ct. at 2159; United States v. Luster, 305 F.3d 199, 202 (3d Cir.2002); United States v. Pierce, 278 F.3d 282, 287 (4th Cir.2002). That is, the offenses defined by those statutes do not have “ \\
& Aldy ex rel. Aldy v. Valmet Paper Mach., 74 F.3d 72, 75 (5th Cir. 1996) (emphasis omitted) (quoting Stena Rederi, 923 F.2d at 386). The third element ensures there is a jurisdictional nexus with the United States. Arriba, 962 F.2d at 533. The “direct effect” requirement involves a determination of whether the acts the suits are “based upon” had “
 \\
& The final element of the “commercial activity” exception poses the issue remaining on this remand, namely, in light of Weltover, did the detention of the aircraft cause “
 \\
\hline

\multirow{3}{*}{Synthetic Naive} 
& What happens when a country refuses to honor an international arbitration award? \\
& Do plaintiffs in ERISA cases involving defined contribution plans have standing to sue for breach of fiduciary duty claims under Section 502(a)(2)? \\
& How do age-related developmental differences impact the evaluation of children in social interactions and behaviors? \\ \hline

\multirow{3}{*}{Synthetic Optimized} 
& waiver of rights by defendant with limited English proficiency \\
& what is the current status of the law on large capacity magazines? \\
& what are the circumstances under which a remand is appropriate? \\

\hline
\multirow{3}{*}{Real-world queries} 
& 803(c)(27) \\
& what is the definition of probable cause? \\
& Is state v. pena-flores still good law? \\
\bottomrule

\end{tabular}

\end{table}

\begin{figure}
    \centering
    \scriptsize
    \begin{tcolorbox}[title=Generating Synthetic Queries]

You are a helpful legal assistant. You are given a paragraph from a legal brief. Your task is to come up with a question / query for which the paragraph would be the top search result. Make sure that the query is not too specific.

Some examples are shown below.

\textit{Examples omitted due to data confidentiality reason}

In case the snippet is not detailed enough or doesn't contain neither facts nor legal reasoning, just reply with None, else return the generated query, nothing else."""
    \end{tcolorbox}
    \caption{System prompt to generate synthetic data. User prompts are all paragraphs in the briefs dataset.}
    \label{fig:system_prompt}
\end{figure}

\begin{figure}
    \centering
\begin{tcolorbox}[title=Anonymizing Paragraphs]
\tiny
You are a PII extraction assistant. You are given a passage from a legal document (a court brief or opinion).

Your task is to identify all personally identifiable information (PII) related to parties in a brief excerpt, and return them as a JSON array of unique strings.

PII to extract contains:
- Names of individuals (parties, judges, attorneys, witnesses)

- Names of specific companies or organizations when they are a party or can identify a party

- Street addresses, cities, ZIP codes when they identify a party

- Docket numbers and case numbers when they identify parties

- Dates of birth, Social Security numbers, phone numbers, email addresses when they identify parties

Do NOT include:

- Generic legal terms, statutory references without jurisdiction identifiers, or procedural terminology

- Legal Citations

- Names that appear within a case citation (e.g., "Smith" in "Smith v. Jones, 123 N.J. 456" should not be extracted)

- Statutory citation abbreviations such as "N.J.S.A." when used as part of a statute reference (e.g., "N.J.S.A. 52:14-7") — these are legal citation shorthand, not identifying references

Return only a JSON array of strings — the unique text spans of each identified PII item — with no explanation, no commentary, and no markdown formatting.
\end{tcolorbox}
    \label{fig:prompt_anonymization}
    \caption{System prompt for anonymizing paragraphs. User prompts are all paragraphs in the dataset.}
\end{figure}

\begin{figure}
    \centering
    \scriptsize
    \begin{tcolorbox}[title=Annotation Guidelines]
Each data point consists of a query submitted by an NJOPD defender and a retrieved paragraph from state directives or briefs. Your task is to annotate  whether the paragraph is useful in public defense work given query. Label each pair as either useful or not useful.

A paragraph is useful if it would plausibly help a public defender who entered that query—for example, if it is topically relevant, cites relevant statutes or precedent, or provides information related to the issue implied by the query. Mentally group each result into no, low, moderate, or high relevance, and mark it as useful if it is moderate or high. If the paragraph is clearly unrelated or unhelpful, label it not useful. For keyword-based queries (e.g., “State v. Pena-Flores”), mark a result as useful if the keyword appears in the paragraph and the content is topically appropriate. If the intent of a query is unclear (e.g., “merger,”), flag the example as unclear intent in a comment.

For each data point, consider the query and retrieved paragraph carefully. You also often see gold paragraphs, examples previously annotated by experienced NJOPD defenders, and any accompanying textual feedback. These should be used as reference points to calibrate your understanding of what constitutes a useful result and to maintain consistency across annotations.

\end{tcolorbox}
    \caption{Annotation guidelines for annotating the PD Dataset.}
\label{fig:annotation_guidelines}
\end{figure}

\begin{figure}
    \centering
    \label{tab:query_expansion_prompt}
    \tiny
    \begin{tcolorbox}[title=System Prompt for IRAC Query Expansion]

Given a legal search query, you should:

1. Infer the key legal issue(s) raised by the query.

2. State the applicable legal rule(s) in general doctrinal terms.

3. Optionally provide a brief legal analysis (reasoning) if it helps clarify the issue and rule.

4. Construct an augmented search query that incorporates the original query plus useful legal reasoning signals (issues, rules, key concepts, doctrinal terms). This augmented query may be in any style (keywords, IRAC-style summary, or a well-structured legal question), as long as it is helpful for retrieving relevant cases and statutes.

Important constraints:

- Do NOT invent or guess specific statute numbers, code sections, or guideline provisions unless they already appear in the original query.

- Do NOT introduce new facts, parties, or jurisdictions that are not clearly implied by the query.

- You MAY generalize to doctrinal labels and concepts (e.g., "merger of offenses", "double jeopardy", "premises liability").

- The augmented query MUST include the original query content in some form (verbatim or lightly edited) plus additional legally-relevant language.

- Be concise but not minimal: prefer adding a few high-value issues/rules/terms over verbose filler.

Output format:

issue: "concise statement of the main legal issue"

rule: "concise statement of the applicable legal rule or doctrine", 

analysis: "optional brief reasoning, 1–3 sentences, may be empty if not helpful",

augmented query: "the final expanded query string used for retrieval"

Remember:

- The `augmented query` can be in natural language or semi-structured (e.g. IRAC-style, dense prose, or keyword-enriched), but it must be optimized to help a legal search engine retrieve the most relevant authorities.

- Do NOT mention statutes or guideline provisions by number if they are not already present in the original query.
    \end{tcolorbox}
    \caption{System prompt for IRAC query expansion. User prompts are all queries in the dataset.}
    
\end{figure}

\end{document}